\documentclass{aa}

\usepackage{epsfig}
\newcommand{\msun}{\mbox{$M_{\odot}$}}
\newcommand{\Msun}{\mbox{$M_{\odot}$}}
\newcommand{\lsun}{\mbox{$L_{\odot}$}}
\newcommand{\Lsun}{\mbox{$L_{\odot}$}}
\newcommand{\rsun}{\mbox{$R_{\odot}$}}

\newcommand{\teff}{\mbox{$T_{\rm eff}$}}
\newcommand{\Teff}{\mbox{$T_{\rm eff}$}}

\newcommand{\vinf}{\mbox{$v_{\infty}$}}
\newcommand{\vesc}{\mbox{$v_{\rm esc}$}}
\newcommand{\ratio}{\mbox{$v_{\infty}$/$v_{\rm esc}$}}
\newcommand{\mdot}{\mbox{$\dot{M}$}}
\newcommand{\Mdot}{\mbox{$\dot{M}$}}

\newcommand{\msunyr}{\mbox{$M_{\odot} {\rm yr}^{-1}$}}

\newcommand{\modfunit}{\mbox{$\msun/{\rm yr}~{\rm km/s}~\rsun^{0.5}$}}

\begin{document}

\thesaurus{07(08.05.1; 08.13.2; 08.19.3; 08.23.3; 08.05.3)}

\title{New theoretical mass-loss rates of O and B stars}

\author{Jorick S. Vink\inst{1}
 \and Alex de Koter\inst{2}
 \and Henny J.G.L.M. Lamers\inst{1,3}}
\offprints{Jorick S. Vink, j.s.vink@astro.uu.nl}

\institute{ Astronomical Institute, Utrecht University,
            P.O.Box 80000, NL-3508 TA Utrecht, The Netherlands
            \and
            Astronomical Institute ``Anton Pannekoek'', University of Amsterdam,
            Kruislaan 403, NL-1098 SJ Amsterdam, The Netherlands
            \and
            SRON Laboratory for Space Research, Sorbonnelaan 2, NL-3584 CA Utrecht, The Netherlands}

\titlerunning{New Mass-loss rates of OB stars}
\authorrunning{Jorick S. Vink et al.}

\maketitle
\begin{abstract}

We have calculated mass-loss rates for a grid of wind models 
covering a wide range of stellar parameters and have derived 
a mass-loss recipe for two ranges of effective temperature at either 
side of the bi-stability jump around spectral type B1. 

For a large sample of O stars, it is shown that there is 
now good agreement between these new theoretical
mass-loss rates that take {\it multiple scattering} into account
and observations.

Agreement between the observed and new theoretical wind momenta
increases confidence in the possibility to derive distances
to luminous stars in distant stellar systems using the 
Wind momentum Luminosity Relation.

For the winds of the B stars there is an inconsistency in the 
literature between various mass-loss rate determinations from 
observations by different methods. One group of $\dot{M}$ determinations 
of B stars {\it does} follow the new theoretical relation, while another 
group does not. The lack of agreement between the observed mass-loss rates 
derived by different methods may point to systematic errors 
in mass-loss determinations from observations for B stars. 

We show that our theoretical mass-loss recipe is reliable and 
recommend it be used in evolutionary calculations.


\end{abstract}


\section{Introduction}
\label{s_intro}

In this paper we present new theoretical mass-loss rates $\dot{M}$
for a wide range of parameters for galactic O and B stars, 
taking {\it multiple scattering} into account. These 
predictions for $\dot{M}$
are compared with observations. The goal of the paper 
is to derive an accurate description of mass loss as a 
function of stellar parameters.

Early-type stars have high mass-loss rates, which
substantially affects their evolution (e.g. Chiosi \& Maeder 1986). 
The winds of early-type stars are most likely 
driven by radiation pressure in lines and 
in the continuum. The radiation-driven wind theory was first developed 
by Lucy \& Solomon (1970) and Castor et al. (1975) (hereafter CAK). 
At a later stage the theory was improved by Abbott (1982), 
Friend \& Abbott (1986), Pauldrach et al. (1986) and 
Kudritzki et al. (1989). 

During the last decade, the radiation-driven wind 
theory has been compared with the most reliable mass-loss 
determinations from observations that are 
available: mass loss determined from radio data and 
from the analysis of H$\alpha$ line profiles. 
Both Lamers \& Leitherer (1993) and Puls et al. (1996) came to the 
conclusion that the theory of radiation-driven winds  
shows a {\it systematic discrepancy} with the observations. 
For O stars the radiation-driven wind theory predicts 
systematically {\it lower} values for mass loss than have been 
derived from observations.

Since the discrepancy increases as a function of wind 
density, it is possible that the reason for this is an 
inadequate treatment of ``multiple scattering'' in the current state of
radiation-driven wind theory. It has been suggested (e.g. by 
Lamers \& Leitherer 1993) that the ``momentum-problem'' that 
has been observed in the dense winds of Wolf-Rayet stars is 
the more extreme appearance of this discrepancy
seen in the winds of normal O-type stars.

Because the observed mass-loss rates for O type supergiants 
are typically a factor of two higher than the values predicted 
by radiation-driven wind theory, evolutionary models would be 
significantly affected if theoretical values were adopted.
It is obvious that an accurate description of mass loss 
is of great importance to construct reliable evolutionary 
tracks for massive stars.

Abbott \& Lucy (1985) and Puls (1987) have investigated 
the importance of ``multiple scattering'' relative to ``single scattering''
for the winds of O stars. Abbott \& Lucy 
found an increase in $\dot{M}$ of a factor of about three for the 
wind of the O supergiant $\zeta$ Puppis if multiple scattering 
was applied in a Monte Carlo simulation. 

We will use a similar Monte Carlo technique in which multiple
scatterings are taken into account to calculate mass-loss rates 
for a wide range of stellar parameters throughout the upper part 
of the Hertzsprung-Russell Diagram (HRD). In Sect.~\ref{s_method}, 
the approach to calculate mass-loss rates will be briefly described,
while in Sect.~\ref{s_massloss}, a grid of wind models and mass-loss 
rates will be presented. A clear separation of the HRD into 
two parts will be made. The first range is that on the ``hot'' 
side of the {\it bi-stability} jump near spectral type B1, 
where the ratio of the terminal velocity to the effective 
escape velocity at the stellar surface ($\ratio$) 
is about 2.6; the second range is that on the ``cool'' side of the jump where 
the ratio suddenly drops to about 1.3 (Lamers et al. 1995). 
At the jump the mass-loss rate is predicted to change dramatically 
due to a drastic change in the ionization of the wind (Vink et al. 1999). 
In Sect.~\ref{s_eta}, the theoretical
wind momentum will be studied and in  
Sect.~\ref{s_recipe} fitting formulae for the mass-loss rate 
will be derived by means of multiple linear regression 
methods: this yields a recipe to predict $\dot{M}$ as a function of 
stellar parameters.
In Sect.~\ref{s_comp} these predicted 
mass-loss rates will be compared with observational rates. 
{\it We will show that for O stars theory and observations agree if ``multiple scattering'' 
is properly taken into account.} Finally, in Sects.~\ref{s_disc} and~\ref{s_concl} 
the study will be discussed and summarized.


\section{Method to calculate \mdot}
\label{s_method}

The basic physical properties of the adopted Monte Carlo (MC)
method to predict mass-loss rates are similar to the 
technique introduced by Abbott \& Lucy (1985).
The precise method was extensively described in Vink et al. (1999).
The core of the approach is that the total loss of radiative
energy is linked to the total gain of momentum of the outflowing material.
The momentum deposition in the wind is calculated by
following the fate of a large number of photons that are
released from below the photosphere. 

The calculation of mass loss by this method 
requires the input of a model atmosphere, before the
radiative acceleration and \mdot\ can be calculated.
The model atmospheres used for this study are calculated
with the most recent version of the non-LTE 
unified\footnote{{\sc isa-wind} 
treats the photosphere and wind in 
a unified manner. This is distinct from the so-called ``core-halo'' approaches.}
Improved Sobolev Approximation code ({\sc isa-wind}) for stars 
with extended atmospheres. For details we refer the reader 
to de Koter et al. (1993,1997). The chemical species that 
are explicitly calculated in non-LTE are H, He, C, N, O 
and Si. The iron-group elements, which are important for 
the radiative acceleration and $\dot{M}$, are treated in the 
modified nebular approximation (see Schmutz 1991).


\section{The predicted mass-loss rates}
\label{s_massloss}

Using the procedure summarized in Sect.~\ref{s_method}, we 
have calculated mass-loss rates for 12 values of $\teff$ in the range
between 12~500 and 50~000 K. For every effective temperature 
a grid of 12 series of models for galactic stars was
calculated with luminosities in the range log ($L_*/\Lsun$) = 4.5 - 6.25 
and masses in the range $M_*$ = 15 - 120  $\Msun$.
For these 144 models, mass-loss rates were calculated for three
values of the ratio $\ratio$, yielding a total number of 432 models.

\begin{table}
\caption[]{Parameters for the 12 ($L_*,M_*$) series of wind models. For 
 details about the model assumptions and grid spacing, see text.}
\label{t_parameters}
\begin{tabular}{rlrcccc}
\hline
\noalign{\smallskip}
series & log$L_{*}$ & $M_*$ & $\Gamma_{\rm e}$ & $M_{\rm eff}$ & \teff & $\left(\frac{\vinf}{\vesc}\right)$\\
\noalign{\smallskip}
no & ($\lsun$) & ($\Msun$) &  & ($\Msun$) & (kK) &  \\
\hline
 1 & 4.5 &  15 & 0.055 & 14.2 & 12.5 - 50.0 &  1.3 - 2.6\\
 2 &     &  20 & 0.041 & 19.2 & 12.5 - 50.0 &  1.3 - 2.6\\
 3 & 5.0 &  20 & 0.130 & 17.4 & 12.5 - 50.0 &  1.3 - 2.6\\
 4 &     &  30 & 0.087 & 27.4 & 12.5 - 50.0 &  1.3 - 2.6\\
 5 &     &  40 & 0.068 & 37.3 & 12.5 - 50.0 &  1.3 - 2.6\\
 6 & 5.5 &  30 & 0.274 & 21.8 & 12.5 - 50.0 &  1.3 - 2.6\\
 7 &     &  40 & 0.206 & 31.8 & 12.5 - 50.0 &  1.3 - 2.6\\
 8 &     &  50 & 0.165 & 41.8 & 12.5 - 50.0 &  1.3 - 2.6\\
 9 &5.75 &  45 & 0.325 & 30.4 & 12.5 - 50.0 &  1.3 - 2.6\\
10 & 6.0 &  60 & 0.434 & 34.0 & 12.5 - 50.0 &  1.3 - 2.6\\
11 &     &  80 & 0.326 & 53.9 & 12.5 - 50.0 &  1.3 - 2.6\\
12 &6.25 & 120 & 0.386 & 73.7 & 12.5 - 50.0 &  1.3 - 2.6\\
\noalign{\smallskip}
\hline
\end{tabular}
\end{table}

\begin{figure}
\centerline{\psfig{file=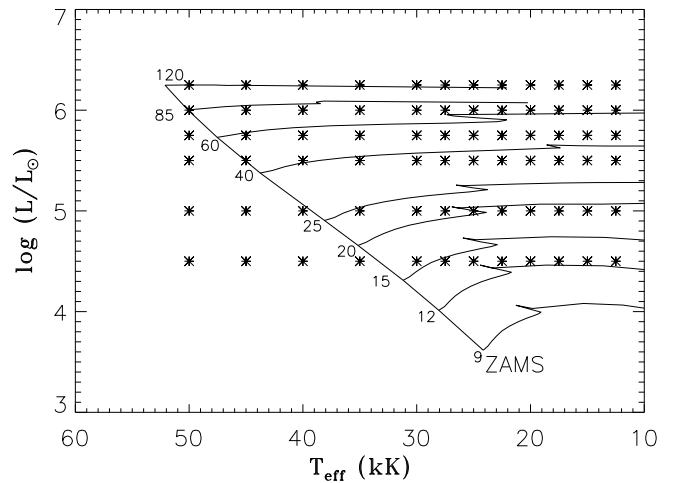, width = 10 cm}}
\caption{Coverage of the calculated wind models over the HRD.
The crosses indicate the model values of log $L/\lsun$ and $\teff$.
Evolutionary tracks from Meynet et al. (1994) are shown for several 
initial masses, which are indicated in the plot. 
The Zero Age Main Sequence (ZAMS) is also shown. 
}
\label{f_HRD}
\end{figure}

The parameters for all series of models are indicated in 
Table~\ref{t_parameters}. In Fig.~\ref{f_HRD} the luminosities and
effective temperatures of the models are indicated with asterisks on top of
evolutionary tracks to show the coverage of the model grid over the 
upper HRD. To study the mass-loss dependence on different stellar 
parameters ($L$, $M$ and $\teff$) separately, a wide range of parameters
was chosen, this implies that some of the models in Fig.~\ref{f_HRD} 
have positions to the {\it left} of the main sequence. 
We enumerate the assumptions in the model grid:

\begin{enumerate}

\item{} The models are calculated for solar metallicities (Allen 1973). 

\item{} The stellar masses in the grid of models were chosen in such a way 
that they are representative for the evolutionary 
luminosities of the tracks from the Geneva group (Meynet et al. 1994). 
To investigate the dependence of $\dot{M}$ on $M_*$, a number 
of smaller and larger values for $M_*$ was also chosen (see column (3) 
in Table~\ref{t_parameters}).

\item{} The grid was constructed in a way that $\Gamma_e \la$ 0.5 
(see column (4) in Table~\ref{t_parameters}), where $\Gamma_e$ is the ratio 
between the gravitational acceleration and the radiative acceleration 
due to electron scattering. $\Gamma_e$ is given by:

\begin{equation}
\label{eq_gammae}
\Gamma_e~=~\frac{L \sigma_e}{4 \pi c G M}~=~7.66~10^{-5} \sigma_{e} 
\left(\frac{L}{\lsun}\right) \left(\frac{M}{\msun}\right)^{-1}
\end{equation}
where $\sigma_e$ is the electron scattering cross-section (its value
is taken as determined in Lamers \& Leitherer 1993) and the  
other constants have their usual meaning.
For values of $\Gamma_{e} > 0.5$, the stars approach their Eddington
limit and the winds show more dramatic mass-loss behaviour. In this 
study, stellar parameters for these ``Luminous Blue Variable-like'' 
stars are excluded to avoid confusion between various physical wind effects.

\item{} All series of models from Table~\ref{t_parameters} have effective temperatures 
between 12~500 and 50~000 K, with a stepsize of 2~500 K from 12~500 to 30~000 K 
and a stepsize of 5~000 K, starting from 30~000 up to 50~000 K.

\item{} We calculated $\dot{M}$ for wind models with a $\beta$-type velocity law
for the accelerating part of the wind:
\begin{equation}
\label{eq_betalaw}
v(r)~=~\vinf~\left(1~-~\frac{R_*}{r}\right)^\beta
\end{equation}
Below the sonic point, a smooth transition from this 
velocity structure is made to a the velocity that follows from the
photospheric density structure.
A value of $\beta=1$ was adopted in the accelerating part of the wind.
This is a typical value for normal OB supergiants 
(see Groenewegen \& Lamers 1989; Haser et al. 1995; Puls et al. 1996).
At a later stage models for other $\beta$ values will be calculated and it 
will be demonstrated that the predicted $\dot{M}$ is essentially 
insensitive to the adopted value of $\beta$ (see Sect.~\ref{s_beta}).

\item{} The dependence of $\mdot$ on various values of $\vinf$ was 
determined.
Lamers et al. (1995) found that the ratio $\ratio \simeq 2.6$ for stars of types
earlier than B1, and drops to $\ratio \simeq 1.3$ for stars later than type B1. 
Therefore, we calculated mass-loss rates for various input values of this ratio,
namely 1.3, 2.0 and 2.6 to investigate the mass loss as a function of this parameter,
similar to that in Vink et al. (1999).
For the determination of $\vesc$, the effective 
mass $M_{\rm eff}=M_*(1-\Gamma_e)$ was used. $M_{\rm eff}$ is given in
column (5) of Table~\ref{t_parameters}.

\end{enumerate}

\begin{table}
\caption[]{Bi-stability jump characteristics for the 12 ($L_*,M_*$) series of wind models.}
\label{t_jumps}
\begin{tabular}{rlrccc}
\hline
\noalign{\smallskip}
series & log $L_*$ & $M_*$ & log($\Delta \dot{M}$) & ${T}_{\rm eff}^{\rm jump}$ & $<\rho>^{\rm jump}$  \\
\noalign{\smallskip}
no & ($\lsun$) & ($\Msun$) &  & (K) & (g cm$^{-3}$)  \\
\hline
 1 &  4.5 &  15 & 0.78 & 23 750 & -14.82  \\
 2 &      &  20 & 0.61 & 22 500 & -15.13  \\
 3 &  5.0 &  20 & 0.83 & 26 250 & -14.22  \\
 4 &      &  30 & 0.87 & 25 000 & -14.68  \\
 5 &      &  40 & 0.73 & 25 000 & -14.74  \\
 6 &  5.5 &  30 & 0.76 & 26 250 & -13.89  \\
 7 &      &  40 & 0.81 & 26 250 & -14.13  \\
 8 &      &  50 & 0.82 & 25 000 & -14.40  \\
 9 & 5.75 &  45 & 0.77 & 25 000 & -13.93  \\
10 &  6.0 &  60 & 0.76 & 25 000 & -13.66  \\
11 &      &  80 & 0.76 & 26 250 & -13.89  \\
12 & 6.25 & 120 & 0.77 & 25 000 & -13.87  \\
\noalign{\smallskip}
\hline
\end{tabular}
\end{table}

\begin{figure*}
\centerline{\psfig{file=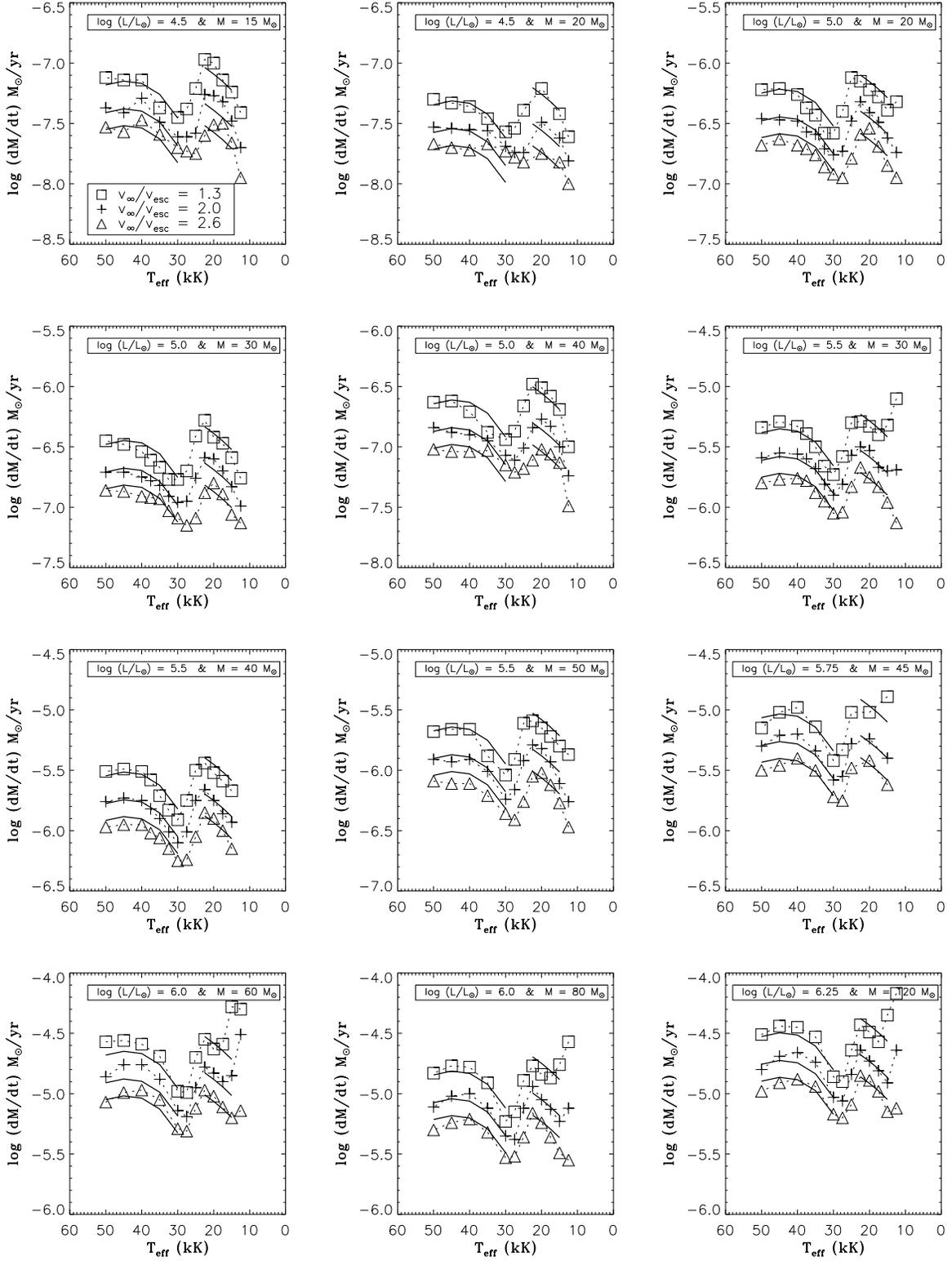, width = 16.0cm}}
\caption{The calculated mass-loss rates $\dot{M}$ as a 
function of $\teff$ for the grid of 12 ($L_*,M_*$) series for three 
values of the ratio $\ratio$. The values for $\ratio$ are indicated 
in the lower part of the first panel. The stellar parameters 
$L_*$ and $M_*$ are indicated in the upper part of each panel. 
The thin dotted lines connect the calculated mass-loss rates. 
The thick solid lines indicate two multiple linear regression 
fits to the calculated values (see Sect.~\ref{s_recipe}).
Note that some of the panels seem to indicate that 
a more accurate fit may be possible. However, the fits have been derived
by {\it multiple} linear regression methods and thus the thick solid lines 
show the mass loss as a function of more than just one parameter. 
All models were calculated for solar metallicities.
}
\label{f_massloss}
\end{figure*}

\subsection{$\dot{M}$ for supergiants in Range 1 \\
~~~~~(30~000 $\le \teff \le$ 50~000 K)}
\label{s_range1}

The results for the complete grid of all the 12 ($L_*,M_*$) series 
are plotted in the individual panels of Fig.~\ref{f_massloss}.
Note that for {\it each} calculated point in the grid, {\it several} 
wind models had to be calculated to check which adopted mass-loss 
rate was consistent with the radiative acceleration 
(see Lucy \& Abbott 1993). This yields predicted, 
self-consistent values for $\mdot$ (see Vink et al. 1999). 

For each ($L_*,M_*$) set and for each value of 
$\ratio$, we found that the mass loss decreases for decreasing  
effective temperature between 50~000 and 27~500 K. 
The reason for this fall-off is essentially that the 
maximum of the flux distribution gradually 
shifts to longer wavelengths. Since there are significantly less 
lines at roughly $\lambda \ga$ 1800 \AA\ than at shorter wavelength, 
the line acceleration becomes less effective at lower 
$T_{\rm eff}$, and thus $\dot{M}$ decreases. 

\subsection{$\dot{M}$ at the bi-stability jump around 25~000 K}
\label{s_jump}

Between about $\teff = 27~500$ and 22~500 K the situation
is reversed: in this range the mass loss {\it increases} drastically 
with decreasing \Teff. These increments in $\dot{M}$ coincide both 
in $\teff$ and in size of the $\dot{M}$ jump with the bi-stability 
jump that was studied by Vink et al. (1999). They showed that the 
origin of the jump is linked to a shift in the ionization balance 
of iron in the lower part of the wind and that it is this element that 
dominates the line driving at the base of the wind.
Below $\teff$ $\simeq$ 25~000 K, Fe {\sc iv} recombines to 
Fe {\sc iii} and as this latter ion is a more efficient line driver 
than Fe {\sc iv}, the line acceleration in the lower part of the 
wind increases. This results in an upward jump in $\dot{M}$ of about a 
factor of five and subsequently a drop in $\vinf$. The drop in $\vinf$ was 
predicted to be a factor of two, which is 
confirmed by determinations of $\vinf$ of OB supergiants 
from ultraviolet data by Lamers et al. (1995). A comparison
between the spectral type of the {\it observed} bi-stability jump and the 
effective temperature of the {\it predicted} jump, was made in 
Vink et al. (1999).

Since we know from both theory and observations 
that the ratio $\ratio$ jumps from $\sim$ 2.6 at the hot side of the jump 
to $\sim$ 1.3 at the cool side of the jump, we can predict the jump in mass loss
for all 12 ($L_*,M_*$) series of models. The size of the jump is defined
as the difference between the minimum $\dot{M}$ at the hot side of the 
jump (where $\ratio$ = 2.6) and the maximum $\dot{M}$ at the cool side (where
$\ratio$ = 1.3) in Fig.~\ref{f_massloss}. The size of the predicted
jump in $\dot{M}$ (log $\Delta\dot{M}$) is indicated in column 
(4) of Table~\ref{t_jumps}: $\Delta\dot{M}$ 
is about a factor of five to seven. Table~\ref{t_jumps} tabulates additional 
characteristics for the models at the bi-stability jump.

The jump in mass loss around $\teff$ $\simeq$ 25~000 K is not exactly the same 
for all series of models: the jump occurs at somewhat different effective  
temperatures. This is no surprise, since the ionization equilibrium 
does not only depend on temperature, but on density as 
well. A smaller value of the ratio
$\ratio$ leads to a larger density in the wind. Hence, the jump is 
expected to start at higher $\teff$ for smaller 
$\ratio$. This behaviour for the position of $\teff$ of the jump is 
confirmed by all individual panels in Fig.~\ref{f_massloss}.
To understand the behaviour of the bi-stability jump as
a function of the other stellar parameters, i.e. $M_*$ and $L_*$, we
will compare the wind characteristics of the 12 series of models 
around the bi-stability jump in some more detail. 

\begin{figure}
\centerline{\psfig{file=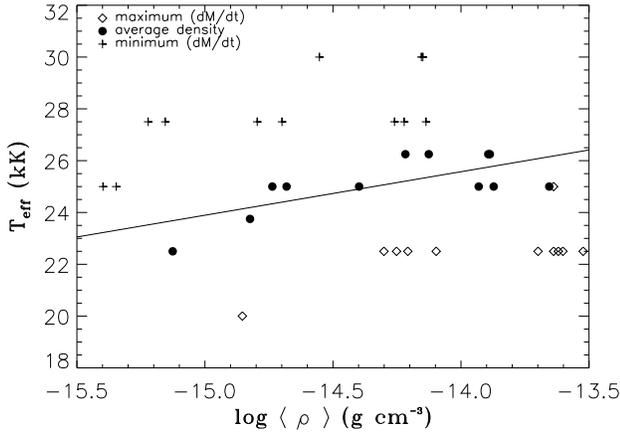, width = 9 cm}}
\caption{Characteristic $<\rho>$ and $\Teff$ of the bi-stability jump around $\teff$ = 25~000 K.
An explanation for the different symbols is given in the legend of the plot. The solid
line represents a linear fit through the average jump parameters log $<\rho>$ and $\teff$.}
\label{f_jump1}
\end{figure}

First we define a characteristic wind density at 50 \% of the
terminal velocity $\vinf$ of the wind: $<\rho>$. For a standard velocity 
law with $\beta=1$, this characteristic wind density is given by

\begin{equation}
<\rho>~=~\frac{\dot{M}}{8 \pi R_*^2 \vinf}
\label{eq_cdens}
\end{equation}
For all 12 series
of models this characteristic density $<\rho>$ is plotted vs. the effective
temperature of the jump. This is done for both the minimum 
$\dot{M}$ (at the hot side of the jump) and the maximum
$\dot{M}$ (at the cool side of the jump).
Figure~\ref{f_jump1} shows the location of the bi-stability jump
in terms of $\teff$ as a function of $<\rho>$.
The characteristic densities and effective temperatures for the cool side 
of the jump are indicated with ``diamond'' signs and with ``plus'' signs for the hot side.
As expected, for all 12 models the minimum $\mdot$ corresponds to a relatively
low $\rho$ and relatively high $\teff$, whereas the maximum $\mdot$ corresponds
to a relatively high $\rho$, but low $\teff$. Note that the effective temperature
at minimum and maximum mass loss is not a very smooth function of wind density. 
This is due to our choice of resolution in effective temperature of the grid. 
We have checked whether the obtained minima and maxima were indeed the extreme
mass-loss values by calculating extra models at intermediate values of 
$\teff$. The minimum and maximum $\dot{M}$ values obtained with the initial grid 
resolution were found to be similar to those determined with a the finer resolution. 
We thus concluded that the initial resolution of the grid was justified.

The ``filled circles'' represent the average values of $T_{\rm eff}$ and $<\rho>$
for the ``jump'' model for each ($L_*,M_*$) series. The ``jump'' model is a 
hypothetical model between the two models where $\dot{M}$ is maximal and minimal. 
The solid line indicates the best linear fit through these averages. 
The relation between the jump temperature (in kK) and log $<\rho>$ is given by:

\begin{equation}
T_{\rm eff}^{\rm jump}~=~49.1~(\pm~9.2)~+~1.67~(\pm~0.64)~{\rm log} < \rho > 
\label{eq_jump1}
\end{equation}
The average temperature and density of the jump are given in columns (5) and (6)
of Table~\ref{t_jumps}. Note that the range in $T_{\rm eff}^{\rm jump}$ is relatively small; all
12 series of models have jump temperatures in the range between 22.5 $\la \teff \la$ 26 kK.

\begin{figure}
\centerline{\psfig{file=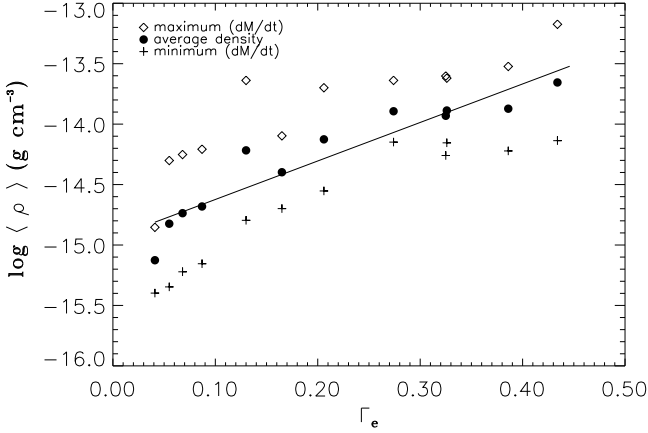, width = 9 cm}}
\caption{Characteristic $<\rho>$ at the bi-stability jump as a function of $\Gamma_e$.
An explanation for the different symbols is given in the legend of the plot. The solid
line indicates a linear fit through the average jump parameters for log $<\rho>$.}
\label{f_gamma1}
\end{figure}

Figure~\ref{f_gamma1} shows the behaviour of the characteristic 
density log $< \rho >$ as a function of $\Gamma_e$. Again this is done for both the 
cool and hot side of the jump, and for the average between them.
As expected, log $<\rho>$ increases as $\Gamma_e$ increases.
Since the average characteristic wind density at the jump
shows an almost linear dependence on $\Gamma_e$, a linear fit through the average
densities is plotted. This is the solid line in Fig.~\ref{f_gamma1}.
The relation between log $<\rho>$ and $\Gamma_e$ is given by:

\begin{equation}
{\rm log} < \rho >~=~- 14.94~(\pm~0.54)~+~3.2~(\pm~2.2)~\Gamma_e
\label{eq_gamma1}
\end{equation}
From the quantities $L_*$ and $M_*$ it is now possible to estimate
log $<\rho>$ using Eq.~(\ref{eq_gamma1}) and subsequently to
predict $T_{\rm eff}^{\rm jump}$ using Eq.~(\ref{eq_jump1}). 
Later on this will be used as a tool to connect two fitting
formulae for the two ranges in $\teff$ at either side 
of the bi-stability jump (see Sect.~\ref{s_recipe}).

\subsection{$\dot{M}$ for supergiants in Range 2 \\
~~~~~(12~500 $\le \teff \le$ 22~500 K)}
\label{s_below225}

Figure~\ref{f_massloss} shows that at 
effective temperatures $\teff \le$ 22~500 K, $\dot{M}$
initially decreases. This is similar to the $\dot{M}$ behaviour 
in the $\teff$ range between 50~000 and 27~500 K.
For some series (dependent on the adopted $L_*/M_*$) the mass loss 
decreases until our calculations end at $\teff$ = 12~500. For other series
of $L_*$ and $M_*$, the initial decrease suddenly
switches to another {\it increase}.
Vink et al. (1999) already anticipated that somewhere, at lower $\teff$, 
a recombination would occur from 
Fe~{\sc iii} to {\sc ii} similar to the recombination from 
Fe~{\sc iv} to {\sc iii} at $\sim$ 25~000 K. 
Lamers et al. (1995) already mentioned 
the possible existence of such a second bi-stability jump around 
\teff\ = 10~000 K from their determinations of $\ratio$, but the observational 
evidence for this second jump is still quite meagre. 

\subsection{$\dot{M}$ at the {\it second} bi-stability jump around 12~500 K}
\label{s_secjump}

To understand the characteristics of the ``second'' 
bi-stability jump as a function of different stellar parameters ($M_*$ and $L_*$), 
we have also studied the models around this second jump in some more detail.

Since our model grid is terminated at 12~500 K, it is not possible to
determine the maximum $\dot{M}$ of the second bi-stability jump in a consistent way,
similar to that of the first jump discussed in Sect.~\ref{s_jump}. 
Thus, it is not possible to determine the exact size of the second jump in $\dot{M}$. 
Neither is it possible to derive an
accurate equation for the position of the second bi-stability jump 
in $\teff$ (as was done in Eq.~\ref{eq_jump1} for the first jump around 25~000 K).
Still, it is useful to determine a rough relationship 
between the position of the second jump in $\teff$ and the average 
log $<\rho>$ by investigating for each model at which temperature the mass-loss rate
still decreases and for which models approaching the second bi-stability jump, the 
mass loss again increases.

The relation found between the temperature of the second
bi-stability jump and log $<\rho>$ is determined by eye and is roughly given by:

\begin{equation}
T^{\rm jump2}~=~100~+~6~{\rm log} < \rho > 
\label{eq_jump2}
\end{equation}
where $T^{\rm jump2}$ is in kK. 
From the quantities $L_*$ and $M_*$ it is again possible to estimate
log $<\rho>$ using Eq.~(\ref{eq_gamma1}) and then to roughly
predict $T^{\rm jump2}$ using Eq.~(\ref{eq_jump2}). 
This formula will be used for our mass-loss recipe at the 
low temperature side (see Sect.~\ref{s_recipe}).


\section{The wind momentum}
\label{s_eta} 

\subsection{The wind efficiency number $\eta$}

\begin{figure*}
\centerline{\psfig{file=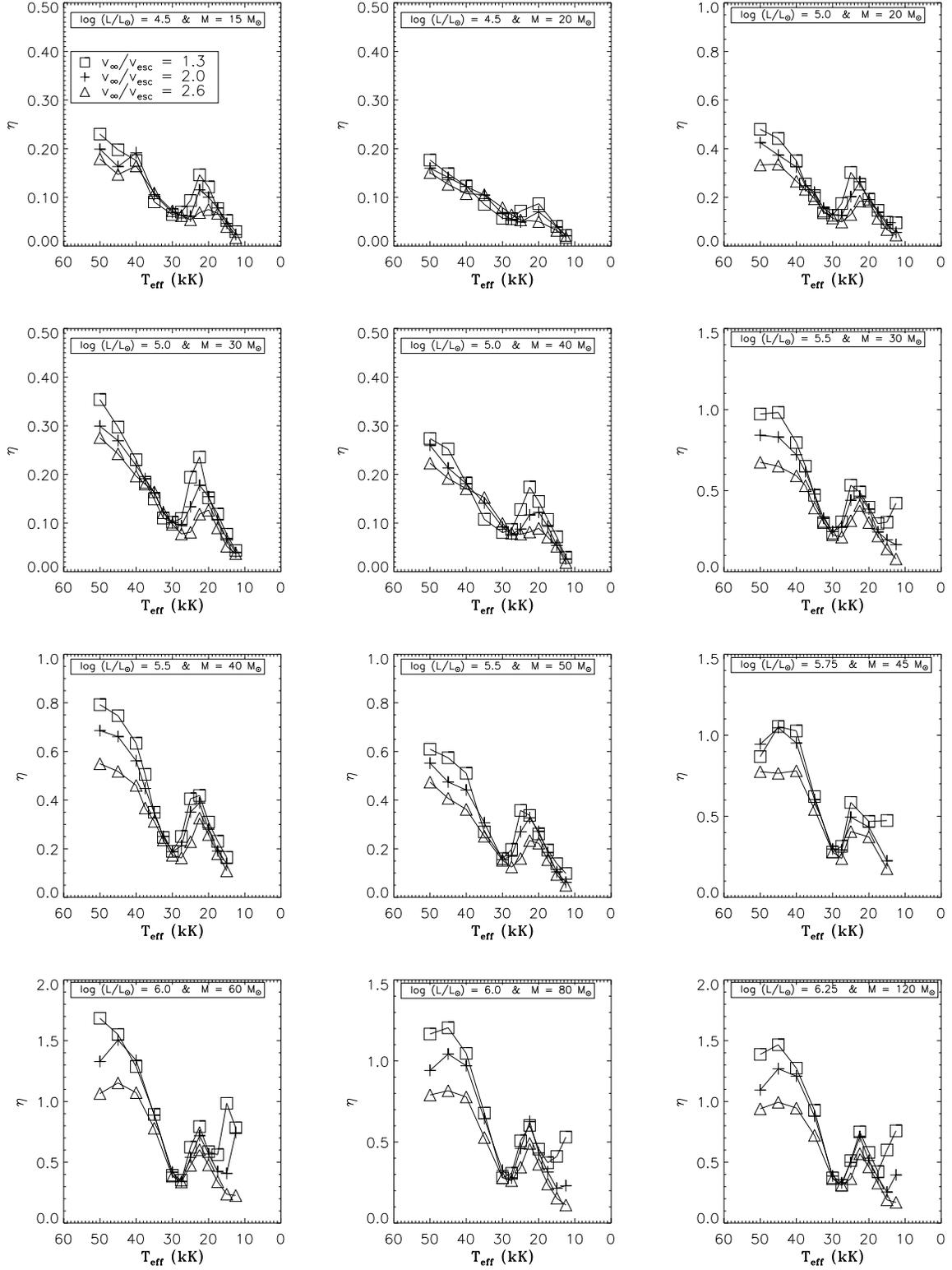, width = 16.0cm}}
\caption{The wind efficiency number $\eta$ as a 
function of $\teff$ for the grid of 12 ($L_*,M_*$) series for three 
values of the ratio $\ratio$. The values for $\ratio$ are indicated 
in the legend of the first panel. The stellar parameters 
are indicated at the top of each panel. All models were calculated for solar 
metallicities.}
\label{f_eta}
\end{figure*}

In this section, we present values for the wind efficiency number $\eta$
for the different $(L_*,M_*)$ series. $\eta$ (sometimes called
the wind performance number) describes the
fraction of the momentum of the radiation that is transferred
to the ions in the wind:

\begin{equation}
\label{eq_eta}
\Mdot \vinf~=~\eta \left( \frac{L_*}{c} \right) 
\end{equation}
Figure \ref{f_eta} shows the behaviour of $\eta$ as a function of $\teff$ for 
the complete grid of models. Figure~\ref{f_eta} demonstrates that {\it $\eta$
is not constant as a function of $\teff$}. The figure shows that when a star
evolves redwards at constant luminosity (from high to low temperature) the momentum
efficiency $\eta$ initially decreases until the star approaches the bi-stability
jump around 25~000 K, where the wind efficiency suddenly {\it increases} 
by a factor of two to three. 
Subsequently, below about 22~500 K, $\eta$ decreases again and in some cases
(again dependent on $L_*$ and $M_*$) it eventually jumps again at 
the second bi-stability jump. This overall behaviour of $\eta$ is 
similar to that of $\dot{M}$ as shown in Fig.~\ref{f_massloss}.

In some of the panels of Fig.~\ref{f_eta}, i.e. in those cases where $L_*/M_*$ is 
large, $\eta$ exceeds the single scattering limit.

\begin{equation}
\eta \equiv \frac{\dot{M} \vinf}{L_*/c} \ge 1
\end{equation}
This occurs at $\teff \ga$ 40~000 K and
log ($L_*/\lsun) \ga$ 6.
It suggests that already for high luminosity OB stars 
stellar winds cannot be treated in the
{\it single scattering} formalism.
The single-scattering limit which is definitely invalid 
for the optically thick winds of Wolf-Rayet type stars, is 
often assumed to be valid for the winds of ``normal'' supergiants. 
Here, however, we come to the conclusion that due to 
{\it multiple scattering}, $\eta$ already exceeds unity for luminous,
but ``normal'' OB supergiants, in case log($L/\lsun) \ga$ 6. This was 
already suggested by Lamers \& Leitherer (1993) on the basis of 
observations.

\subsection{The importance of multiple scattering}
\label{s_importance}

Puls et al. (1996) proposed that the reason for the systematic 
discrepancy between the observed mass-loss rates and  
recent standard radiation driven wind models (Pauldrach et al. 1994) 
was caused by an inadequate treatment of multi-line effects in 
these wind models. To compare our new mass-loss
predictions with the most sophisticated prior investigations, it is 
useful to briefly discuss the most important assumptions that are 
made in modelling the wind dynamics of OB-type stars. 
The following four basic choices must be made:

\begin{enumerate}
\item{} One may treat the photosphere and wind in a ``core-halo''
        approximation, or one may not make this distinction and
        treat photosphere and wind in a ``unified'' way. This 
        choice must be made twice, i.e. with respect to the calculation of the 
        occupation numbers as well as with respect to the computation 
        of the line force.
\item{} One may adopt a ``single-line'' approach, i.e. neglecting 
        effects caused by overlapping lines, or one may follow an 
        approach including ``multi-line'' effects.
\item{} One solves the rate equations for all relevant 
        ions explicitly in non-LTE, or one adopts a ``nebular type of
        approach'' to calculate the ionization balance.
\item{} One solves the equation of motion self-consistently, 
        or one derives the wind properties from a global energy argument.
\end{enumerate}

Standard radiation driven wind models (CAK, Abbott 1982, Pauldrach
et al. 1994) treat the momentum equation in a core-halo approach (1)
adopting the single-line approximation (2). Various degrees of sophistication
can be applied to determine the occupation numbers. The studies
of Pauldrach et al. (1994) and Taresch et al. (1997) represent
the current state-of-the-art, i.e. they treat all relevant ions explicitly 
in non-LTE (3) and solve the equation of motion self-consistently (4).
Pauldrach et al. (1994) also use a unified method for the calculation
of the occupation numbers, but a ``core-halo''approach is applied
with respect to the line force.  Additionally, as line overlap 
is neglected in the method used by Pauldrach et al. (1994), these models 
can overestimate the line force as unattenuated photospheric flux is 
offered to each line, which consequently may produce efficiency numbers 
larger than unity. 
 
Puls (1987) found that for winds of relatively low density (say 
$\eta \la 1/2$) the inclusion of multi-line effects leads to a 
reduction of wind momentum {\em compared to the standard model} due to 
backscattering and blocking of photons in the lower part of the wind. 
For winds of relatively high density (say $\eta \ga 1$), such as the 
dense winds of Wolf-Rayet stars, the situation is likely to be reversed. 
Here momentum transfer from an extended diffuse field is expected to 
dominate over the effect of the attenuation of flux in the layers 
just above the photosphere. This could result in more mass loss
compared to the standard radiation driven wind theory (Abbott \& Lucy 1985, 
Springmann 1994). Wolf-Rayet and Of/WN stars profit from a layered 
ionization structure, which increases the number of lines that can
be used for the driving and thus increasing the mass loss 
(Lucy \& Abbott 1993, de Koter et al. 1997).

Our method differs in almost all aspects from that of 
Pauldrach et al. (1994). In our method, photosphere and wind
are treated in a unified manner (1) and we properly take multi-scatterings
into account with a Monte Carlo technique (2). On the other hand, we 
derive the level populations of the iron-group elements using (a 
sophisticated version of) the nebular approximation (3). Finally,
we derive the mass loss from a global energy argument (4).
This distinct difference of approach implies that a comparison
between both methods is difficult. 
Still, we will address some of the differences in approach by 
focusing on a star with parameters 
representative for the O4I(f)-star $\zeta$~Puppis, which has been 
studied in detail by Abbott \& Lucy (1985), Puls (1987), Pauldrach et al. (1994) 
and Puls et al. (1996).

\begin{figure}
\centerline{\psfig{file=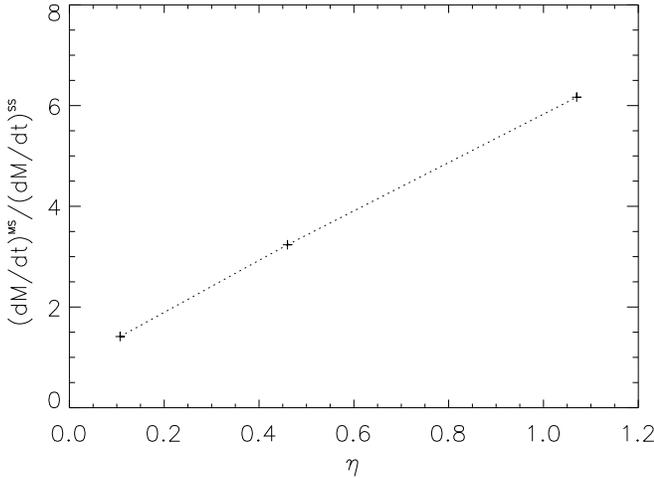, width = 10 cm}}
\caption{The efficiency of multiple-scattering for a range of wind densities.
MS refers to multiple-scattering, and SS refers to single-scattering.}
\label{f_msplot}
\end{figure}

We can test the difference between single scattering and multiple 
scattering by allowing photons to interact with a line only once. 
Fig.~\ref{f_msplot} show a comparison between the single- and 
multiple scattering case for three representative wind 
models at \teff\ = 40,000~K. The model parameters are given in Table~3.
For the often studied wind of the O supergiant $\zeta$ Puppis, 
which has a mass-loss rate of $\dot{M}^{\rm obs} = 5.9 \times 10^{-6} \msunyr$ 
(Puls et al. 1996), the observed efficiency number is about $\eta \simeq 0.6$, 
suggesting that the {\it real} efficiency of multiple vs. single scattering 
is a factor of about four for $\zeta$ Puppis (see Fig.~\ref{f_msplot}).
This is close to the findings of Abbott \& Lucy (1985)
who found an increase in $\dot{M}$ by a factor of 3.3 for the 
wind of $\zeta$ Puppis if multiple scattering was taken into account 
in a Monte Carlo simulation.

Note from the figure that at low wind densities, the single- and 
multiple scattering approach converge, as one would expect. For typical 
O-stars, which have $\eta \la 0.5$, the mass loss will increase by up to 
a factor of two when multiple scattering is properly included. 
The Wolf-Rayet stars, located at the extreme high wind density side, and which 
in some cases have observed efficiency numbers of factors 10 or even higher, may 
benefit by factors of up to $\sim$ 50.  

The reason why Puls (1987) found a reduced mass loss for $\zeta$~Puppis 
when comparing the single-line approach with the multi-line approach is 
because the single-line approach (which is {\em not} the same as the 
single scattering process) overestimates the line force at the base of 
the wind, where the mass loss is fixed. However, a similar relative 
behaviour is {\it not} found when we compare the predicted single-line 
mass loss $\mdot\ = 5.1 \times 10^{-6} \msunyr$ of Pauldrach et al. (1994) 
with the value of $\mdot\ = 8.6 \times 10^{-6} \msunyr$ derived
from our fitting formula based on multiple scattering models. It is not
possible to exactly pinpoint the cause of this difference, but it is likely
to be related to differences in our multi-line treatment and that of Puls (1987).
Contrary to Puls (and also contrary to Abbott \& Lucy 1985), we do not 
adopt the core-halo approximation. The formation region of the strong 
driving lines extends from the photosphere out to the base of the wind. 
If one assumes an input photospheric spectrum representative
of the emergent ultraviolet spectrum as in a core-halo approach, one may 
overestimate the blocking in the subsonic wind regime which results in 
a lower mass loss.

\begin{table}
\caption[]{The relative importance of multiple (MS) vs. single scattering (SS) for a wind model 
at $\teff = 40~000$ K.}
\label{t_importance}
\begin{tabular}{ccccccc}
\hline
\noalign{\smallskip}
$\Gamma_{\rm e}$ & log $L_*$ & $M_*$ & $\eta^{\rm MS}$ & log $\dot{M}^{\rm SS}$ & log $\dot{M}^{\rm MS}$ & $\frac{\dot{M}^{\rm MS}}{\dot{M}^{\rm SS}}$ \\
\noalign{\smallskip}
  & ($\lsun$) & ($\Msun$) &  & & & \\
\hline
0.041  & 4.5 & 20 & 0.107 & -7.87  & -7.72 & 1.41 \\
0.206  & 5.5 & 40 & 0.460 & -6.46  & -5.95 & 3.24 \\
0.434  & 6.0 & 60 & 1.07  & -5.76  & -4.97 & 6.17 \\
\noalign{\smallskip}
\hline
\end{tabular}
\end{table}

\subsection{The Modified Wind Momentum $\Pi$}
\label{s_mwm}

Kudritzki et al. (1995) have defined the Wind 
momentum Luminosity Relation (WLR):

\begin{equation}
\Pi~\equiv~\dot{M}~\vinf~R_*^{0.5}~\propto~L_{*}^{x}
\label{eq_wlr}
\end{equation}
where $\Pi$~(or $\dot{M} \vinf R_*^{0.5}$) is called the ``modified
wind momentum''. Observations of $\dot{M}$ and $\vinf$ of O supergiants have
shown that log $\Pi$ is proportional to log $L_*$ 
(see e.g. Puls et al. 1996). The WLR may in principle be used 
as a tool to derive distances to galaxies (see Kudritzki et al. 1995).

In the theory of line driven winds, the reciprocal value of $x$ equals
(Puls et al. 1996):
\begin{equation}
1/x = \alpha_{\rm eff} = \alpha - \delta
\label{eq_alphaeff}
\end{equation} 
Here $\alpha$ and $\delta$ are force multiplier parameters, 
describing the radiative line acceleration $g_{\rm line}$ 
through the stellar wind:

\begin{equation}
g_{\rm line}~\propto~\left(\frac{1}{\rho} \frac{dv}{dr}\right)^\alpha \left(\frac{n_{\rm e}}{W}\right)^\delta
\label{eq_gline}
\end{equation}
where $n_{\rm e}$ is the electron density and $W$ is the dilution factor.
$\alpha$ corresponds to the power law exponent of the line strength 
distribution function controlling the relative number of strong to weak 
lines. If only strong (weak) lines contribute to the line acceleration
force, then $\alpha$ = 1 (0). The predicted value of $\alpha$ is about 0.6. 
The parameter $\delta$ describes the ionization balance of the wind. 
Values for this parameter are usually between 0.0 and 0.1. 
For a detailed discussion of the parameterisation of the line
acceleration, see e.g. CAK, Abbott (1982) and Kudritzki et al. (1989).

The important point to note here is that possible changes in the slope $x$
as a function of effective temperature reflect the fact that the stellar 
winds are driven by different sets of ions, i.e. lines of different ions.
Figure~\ref{f_eta} shows that around the bi-stability jump at 
$\teff$ $\simeq$ 25~000 K, $\eta$ {\it increases} for 
decreasing $\teff$. This implies that one does not necessarily expect 
a universal WLR over the complete spectral range of O, B and A stars,
nor does one expect a constant value of $\alpha_{\rm eff}$ or $x$
for different spectral types.


\section{Mass loss recipe}
\label{s_recipe}

In this section we present a theoretical mass loss formula for OB stars over
the full range in $\teff$ between 50~000 and 12~500 K. The mass-loss rate
as a function of four basic parameters will be provided. These parameters
are the stellar mass and luminosity, effective temperature and terminal velocity 
of the wind.
To obtain a mass-loss recipe, we have derived interpolation 
formulae from the grid of $\mdot$ calculations 
presented in Sect.~\ref{s_massloss}. The fitting procedure
was performed using multiple linear regression methods to derive
dependence coefficients. We have 
applied this method for the two ranges in $\teff$ separately. The first range is roughly the range
for the O-type stars between $\teff$ = 50~000 and 30~000 K. The second range 
is between $\teff$ = 22~500 and 15~000 K, which is roughly the range
for the B-type supergiants. The two relations are connected at the bi-stability jump.
We have already derived the jump parameters for different series
of models in Sect.~\ref{s_massloss}, so we have knowledge about 
the position of the jump in $\teff$ as a function of stellar parameters. 
This will be applied in the determination of mass loss for stars with temperatures
around the bi-stability jump.

\subsection{Range 1 (30~000 $\le \teff \le$ 50~000 K)}
\label{s_Ofit}

The first range (roughly the range of the O-type stars) is taken from
$\teff$ between 50~000 K and 30~000 K. 
In this range the step size in effective temperature 
equals 5~000 K. So, for the first range 
we have five grid points in $\teff$. Five times 12 series of ($L_*,M_*$),
together with three ratios of ($\ratio$) yields a total of 180 points in $\dot{M}$ 
for the first range.
We have found that for the dependence of $\dot{M}$ on $\teff$, the 
fit improved if a second order term (log $\teff)^{2}$ was taken 
into account. In fact, this is obvious 
from the shapes of the plots in the panels of Fig.~\ref{f_massloss}. 
The best fit that was found by multiple linear regression is:

\begin{eqnarray}
{\rm log}~\dot{M} & = &~-~6.697~(\pm 0.061) \nonumber \\
                  & &~+~2.194~(\pm 0.021)~{\rm log}(L_*/{10^5}) \nonumber \\
                  & &~-~1.313~(\pm 0.046)~{\rm log}(M_*/30) \nonumber\\
                  & &~-~1.226~(\pm 0.037)~{\rm log}\left(\frac{\ratio}{2.0}\right) \nonumber \\
                  & &~+~0.933~(\pm 0.064)~{\rm log}(\teff/40 000) \nonumber\\
                  & &~-~10.92~(\pm 0.90)~\{{\rm log}(\teff/40 000)\}^{2} \nonumber\\
                  \nonumber\\
                  & &~{\rm for}~27~500 < \teff \le 50~000 {\rm K}
\label{eq_Ofit}
\end{eqnarray}
where $\dot{M}$ is in $\msun$ ${\rm yr}^{-1}$, $L_*$ and $M_*$ are in solar units
and $\teff$ is in Kelvin. Note that $M_*$ is the stellar mass 
{\it not} corrected for electron scattering. In this range $\ratio$ = 2.6. 
Equation ~\ref{eq_Ofit} predicts the calculated mass-loss rates of the 
180 models with a root-mean-square (rms) accuracy of 0.061 dex.
The fits for the various ($L_*,M_*$)
series are indicated with the thick lines in the panels of Fig.~\ref{f_massloss}.
Note that some of the panels in Fig.~\ref{f_massloss} seem to indicate that 
a more accurate fit might have been possible. However, Eq.~(\ref{eq_Ofit}) is derived
by {\it multiple} linear regression methods and thus it provides
the mass loss as a function of more than just one parameter.

\subsection{Range 2 (15~000 $\le \teff \le$ 22~500 K)}
\label{s_Bfit}

The second range (roughly the range of the B-type supergiants) is taken from
$\teff$ between 22~500 and 15~000 K. 
In this range the step size in effective temperature 
equals 2~500 K. For this range, there are 
four grid points in $\teff$. Four times 12 series of ($L_*,M_*$),
together with three ratios of ($\ratio$) yields a total of 144 points in $\dot{M}$.
In this range the fit did not improve if a second 
order term in effective temperature was taken into account, so this was
not done.
The best fit that was found by multiple linear regression for the second range is:

\begin{eqnarray}
{\rm log}~\dot{M} & = &~-~6.688~(\pm 0.080) \nonumber \\
                  & &~+~2.210~(\pm 0.031)~{\rm log}(L_*/{10^5}) \nonumber \\
                  & &~-~1.339~(\pm 0.068)~{\rm log}(M_*/30) \nonumber\\
                  & &~-~1.601~(\pm 0.055)~{\rm log}\left(\frac{\ratio}{2.0}\right) \nonumber \\
                  & &~+~1.07~(\pm 0.10)~{\rm log}(\teff/20 000) \nonumber\\
                  \nonumber\\
                  & &~{\rm for}~12~500 < \teff \le 22~500 {\rm K}
\label{eq_Bfit}                  
\end{eqnarray}
where again $\dot{M}$ is in $\msun$ ${\rm yr}^{-1}$, $L_*$ and $M_*$ 
are in solar units and $\teff$ is in Kelvin. In this range $\ratio$ = 1.3. 
The fitting formula is also indicated by solid lines in 
the panels of Fig.~\ref{f_massloss}. Equation~\ref{eq_Bfit} predicts 
the calculated mass-loss rates of the 144 models for this $\teff$ range 
with an rms accuracy of 0.080 dex. For this second range 
(12~500 $< \teff \le$ 22~500~K) the fit is slightly less good than for the 
first $\teff$ range. This is due to
the presence of the second bi-stability jump which already appears
in some ($L_*/M_*$) cases, as was shown in Fig.~\ref{f_massloss}.
If those models that do show the {\it second} bi-stability jump, i.e.
stars with high $\Gamma_e$, are omitted from the sample, the accuracy
improves to $\simeq 0.06$ dex.  
In all cases the rms is $\la$ 0.08 dex in log $\dot{M}$, which
implies that the fitting formulae yield good representations of 
the actual model calculations. 

We are aware of the fact that there could be systematic errors in our
approach, since we have made assumptions in our modelling. For a
discussion of these assumptions, see Vink et al. (1999). 
Whether there are still systematic errors between the observed 
mass-loss rates and these new predictions of radiation-driven wind theory, 
will be investigated in Sect.~\ref{s_comp}. 

\subsection{The complete mass-loss recipe}

For stars with effective temperatures higher than 27~500 K, one should apply
the mass-loss formula for the first range (Eq.~\ref{eq_Ofit}); for 
stars with $\teff$ lower than 22~500 K the formula 
for the second range (Eq.~\ref{eq_Bfit}) is to be used. 
In the range between 22~500 and 27~500 K, it is not a priori known 
which formula to apply. This due to the presence of the bi-stability jump. 
Nevertheless, it is possible to retrieve a reliable mass-loss prediction
by using Eqs.~\ref{eq_jump1} and~\ref{eq_gamma1} as a tool to determine
the position of the jump in $\Teff$.

In predicting the mass-loss rate of stars close to the bi-stability
jump, one should preferentially use the {\it observed} 
$\ratio$ value to determine the position with respect to 
the jump. This is a better approach than to use the tools from 
Eqs.~\ref{eq_jump1} and \ref{eq_gamma1} to determine
the position of the jump. The reason is that errors in the basic stellar 
parameters may accidently place the star at the wrong side of the jump. 
A computer routine to calculate mass loss as a function of input
parameters is available either upon request or at the following url:
www.astro.uu.nl/$\sim$jvink/. 

If $\vinf$ is not available, as is the case for 
evolutionary calculations, one should adopt the ratio $\ratio$ = 2.6 
for the hot side of the jump and $\ratio$ = 1.3 for the cool side of the jump, 
in agreement with the analysis by Lamers et al. (1995). 
Note that the exact $\teff$ of the jump is not expected to have 
a significant effect on evolutionary tracks calculated with this new mass-loss
description, 
since the most luminous stars spend 
only a relatively short time around $T_{\rm eff} \simeq$ 25~000 K during their evolution.

Since our calculations were terminated at 12~500 K, we are not able
to determine the size and the position of the second bi-stability jump.
Predicting the mass-loss behaviour below this second jump would therefore be 
speculative.
Yet, for evolutionary tracks the mass loss below 
12~500 K is an important ingredient in the evolutionary 
calculations. We roughly estimate from our grid calculations 
that for a {\it constant} ratio of $\ratio$ the increase
in $\dot{M}$ around 12~500 is about a factor of two, similar to that found 
for the first jump near 25~000 K. Furthermore, observations by Lamers et al. (1995) 
indicate that for stars around 10~000 K, $\ratio$ drops again by a factor of 
two from $\ratio \simeq$ 1.3 to about 0.7. It is therefore plausible to 
expect that the size in $\dot{M}$ of the second jump is 
comparable to the size of the first jump. So, $\Delta \dot{M}$ of the 
second jump is expected to be a factor of five also. 
We argue that this second jump should also be 
considered in evolutionary calculations and suggest Eq.~(\ref{eq_Bfit})
could be used for effective temperatures below the second jump when 
the constant in Eq.~(\ref{eq_Bfit}) is increased by a factor of 
five (or log $\Delta \dot{M}$ = 0.70) to a value of -5.99. 
The mass-loss recipe can be applied for evolutionary calculations 
until the point in the HRD where line driven winds become inefficient and 
where probably another mass-loss mechanism switches on for the cooler
supergiants (see Achmad et al. 1997). We suggest that 
in the temperature range below the second jump $\ratio$ = 0.7 is adopted.

\subsection{The dependence of $\dot{M}$ on the steepness of the velocity law $\beta$}
\label{s_beta}

To test the sensitivity of our predictions of mass-loss rates on different
shapes of the velocity law, we have calculated series of models for
$\beta$ = 0.7, 1.0 and 1.5. This is a reasonable range 
for OB stars, see Groenewegen \& Lamers 1989; Puls et al. 1996). 
The adopted stellar parameters for this test are 
$L_*=10^5$ \Lsun\ and $M_* = 20~\Msun$. We have calculated 
$\dot{M}$ for all of the above $\beta$ values for wind models with 
the three values $\ratio$ = 2.6, 2.0 and 1.3.

\begin{figure}
\centerline{\psfig{file=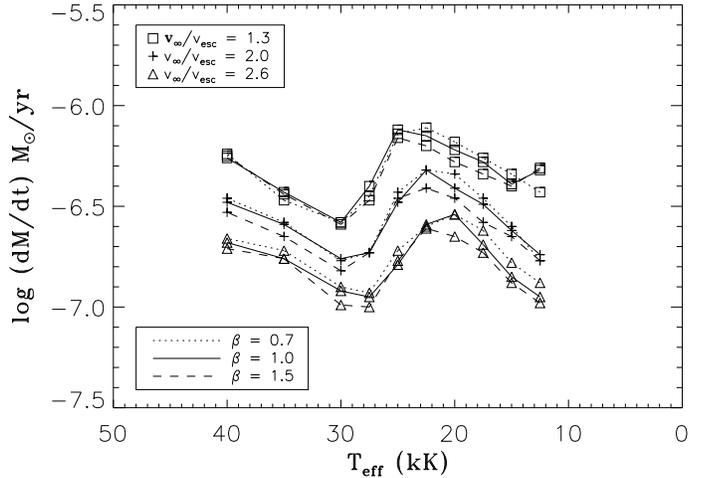, width = 10 cm}}
\caption{Dependence of $\dot{M}$ on the shape of the velocity law, 
for three values of $\beta$ = 0.7, 1.0 and 1.5, as is 
indicated in the lower left corner of the plot.
The values for $\ratio$ are indicated in the upper left corner of the plot. 
For other stellar parameters, see text.}
\label{f_beta}
\end{figure}

From the results shown in Fig.~\ref{f_beta} we derived for the
dependence of $\dot{M}$ on the adopted value of $\beta$:

\begin{equation}
{\rm log}~\dot{M} = C~+~0.112~(\pm 0.048)~{\rm log}(\beta/{1.0}) 
\label{eq_beta}                  
\end{equation}
where $C$ is a constant. This relation is valid for the 
range between $\beta$ = 0.7 - 1.5.
Since the dependence on this parameter is significantly smaller than that
on the other parameters, $L_*$, $M_*$, $\teff$ and $\ratio$,
as was found in Eqs.~\ref{eq_Ofit} and~\ref{eq_Bfit}, we have omitted
the $\beta$ dependence from the mass loss recipe. 
We have just presented the $\beta$ dependence in this section for the sake
of completeness, but we can conclude that the predicted mass-loss rates are 
only marginally sensitive to the {\it shape} of the adopted velocity law. 
One could argue that a $\beta$ dependence on $\dot{M}$ could be of significance
for more extreme series of models. This was 
tested, but it turned out that for the high $\Gamma_e$ series,
the $\beta$ dependence is also insignificant, i.e. deviations of predicted
$\dot{M}$ are less than $\Delta \dot{M} \la 0.03$ dex.
This shows that we can safely omit the $\beta$ dependence on $\dot{M}$ in the
mass-loss recipe for the O and B stars.

\begin{figure*}
\centerline{\psfig{file=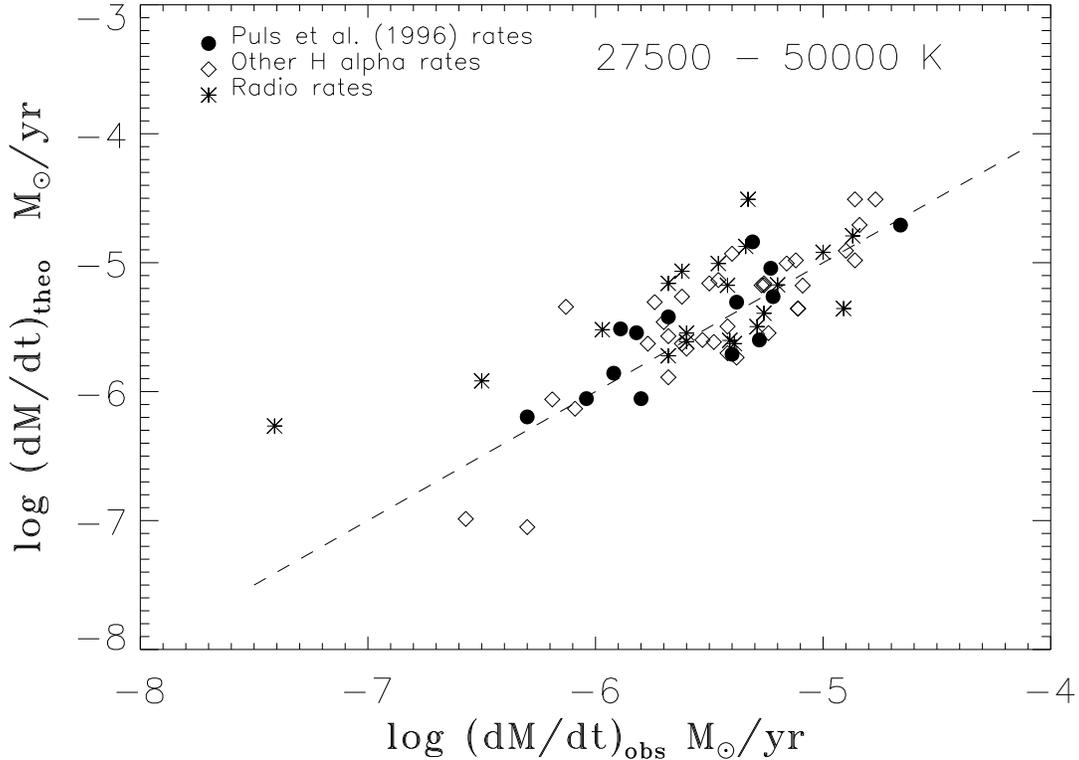, width = 16cm}}
\caption{Comparison between theoretical and observational $\dot{M}$ 
        (both radio data and H$\alpha$) for the O stars.
        The Puls et al. (1996) H$\alpha$ rates; H$\alpha$ rates from other
        determinations, and radio mass-loss rates are indicated with different symbols.
        The dashed line is a one-to-one relation.}
\label{f_mdotO}
\end{figure*}


\section{Comparison between theoretical and observational $\dot{M}$}
\label{s_comp}

\subsection{$\dot{M}$ comparison for Range 1 \\
~~~~~(27~500 $< \teff \le$ 50~000 K)}
\label{s_masslosscomp}

An extended compilation of observed mass-loss rates of 
early-type OBA stars is obtained by Lamers et al. (2000; in preparation).
Since both the ultraviolet and the infrared method do not yet 
yield reliable rates, only mass-loss rates based on radio free-free 
emission and emission of H$\alpha$
have been considered. The H$\alpha$ 
mass-loss rates and their stellar parameters are 
from: Herrero et al. (2000); Kudritzki et al. (1999);
Lamers \& Leitherer (1993) (these H$\alpha$ equivalent width values
are corrected with the curve of growth method from Puls et
al. 1996); Puls et al. (1996); Scuderi et al. (1992), Scuderi (1994),
and Scuderi \& Panagia (2000).
The radio rates are from the compilation of Lamers \& Leitherer (1993);
from Leitherer et al. (1995) and Scuderi et al. (1998). The observed terminal
velocities are from the same papers. These were mainly determined from P Cygni 
profiles. The stellar masses are derived from evolutionary tracks of 
Meynet et al. (1994). For a critical discussion of the
observed mass-loss rates and for the selection of the most reliable data, see
Lamers et al. (in preparation). 

For all these stars with known observational mass-loss rates and stellar
parameters, we have determined theoretical $\dot{M}$ values with the mass-loss recipe 
that was derived in Sect.~\ref{s_recipe}.
A star-to-star-comparison between these predicted mass-loss rates and 
those derived from observations is presented in Fig.~\ref{f_mdotO}.
In this plot only the stars above the bi-stability jump 
(where $\teff \ge$ 27~500 K) are included.
The mass-loss rates from Puls et al. (1996) are indicated with a different 
symbol (filled circle), because these are obtained from a homogeneous set, 
and are analyzed with the most sophisticated wind models. 
Note that the outlier at 
log $\dot{M}_{\rm obs}$ $\simeq$ -~7.4 is the star 
$\zeta$ Oph (HD 149757) for which Lamers \& Leitherer (1993) 
reported that 
the mass-loss rate is uncertain.

The errors in Fig.~\ref{f_mdotO} can be due to several effects. There is an 
error in the theoretical fitting formula, though this error is 
only 0.061 dex (see Sect.~\ref{s_Ofit}). There could also be systematic 
errors due to assumptions in the modelling.
Furthermore, there could be systematic errors in the mass-loss determinations 
from observations. Such systematic effects may for instance occur if the 
clumping factor in the wind changes with distance to the central star. 
This because the H$\alpha$ and radio
emission originate from distinctly different regions in the stellar wind.
However, Lamers \& Leitherer (1993) have shown that for a significant
sample of O stars there is good 
agreement between the radio and the H$\alpha$ mass-loss rates. 
 
The random errors in the observational mass-loss rates are due to 
uncertainties in the stellar parameters and in the mass-loss determinations. 
We tentatively estimate the intrinsic errors in the observed mass-loss rates 
from the radio and H$\alpha$ method to be on the order of 0.2 - 0.3 dex 
(see Lamers et al. in preparation).
This means that for a star-by-star comparison between observations and theory 
one would expect a scatter around the mean which is a combination of the theoretical
and observed uncertainties. 
This error is on the order of 0.3 dex. 
The scatter between observational and theoretical mass-loss rates 
for the O stars from Fig.~\ref{f_mdotO} that was actually derived,  
equals 0.33 dex ( 1 $\sigma$) for the complete set and is 0.24 dex for the 
Puls et al. set. This is an expected scatter and it implies that 
{\it we do not find a systematic discrepancy between observations and our predictions 
for the O star mass-loss rates}.

Contrary to earlier comparisons between observations and theory 
where systematic discrepancies have
been reported (see Lamers \& Leitherer 1993, Puls et al. 1996), here we find that 
there is agreement between our predictions and the mass-loss 
rates derived from observations for the O-type stars. 
The essential difference between previous studies and the present
one is that in our treatment of the theory of line driven winds, we 
consistently take into account effects of ``multiple-scattering''
in the transfer of momentum from the radiation field to the wind. 
We find systematic agreement
between observed and theoretical mass-loss rates for a {\it large sample} of
O stars. This result implies that physical effects that were not incorporated 
in our models, such as magnetic fields and stellar rotation, is not 
expected to influence the mass-loss rates of O stars significantly.


\subsection{Modified Wind momentum comparison for Range 1 \\
~~~~~(27~500 $< \teff \le$ 50~000 K)}
\label{s_modfcompO}

Instead of comparing just the mass-loss rates it is useful to compare 
(modified) wind momenta derived from observations and theory. 
In earlier studies, e.g. Lamers \& Leitherer (1993), and Puls et al. (1996), 
wind momenta have been plotted versus the wind efficiency number $\eta$. 
Comparisons between observed and theoretical wind momenta as a function
of $\eta$ could yield important information about the origin of the
systematic discrepancy between theory and observations.
However, since these two quantities (wind momentum and
wind efficiency number) both contain the mass-loss rate, they 
are not independent. Therefore, no such comparison is made here. 
Instead, the wind momenta are plotted versus the stellar luminosity,
to compare the observational and theoretical WLR. 

\begin{figure}
\centerline{\psfig{file=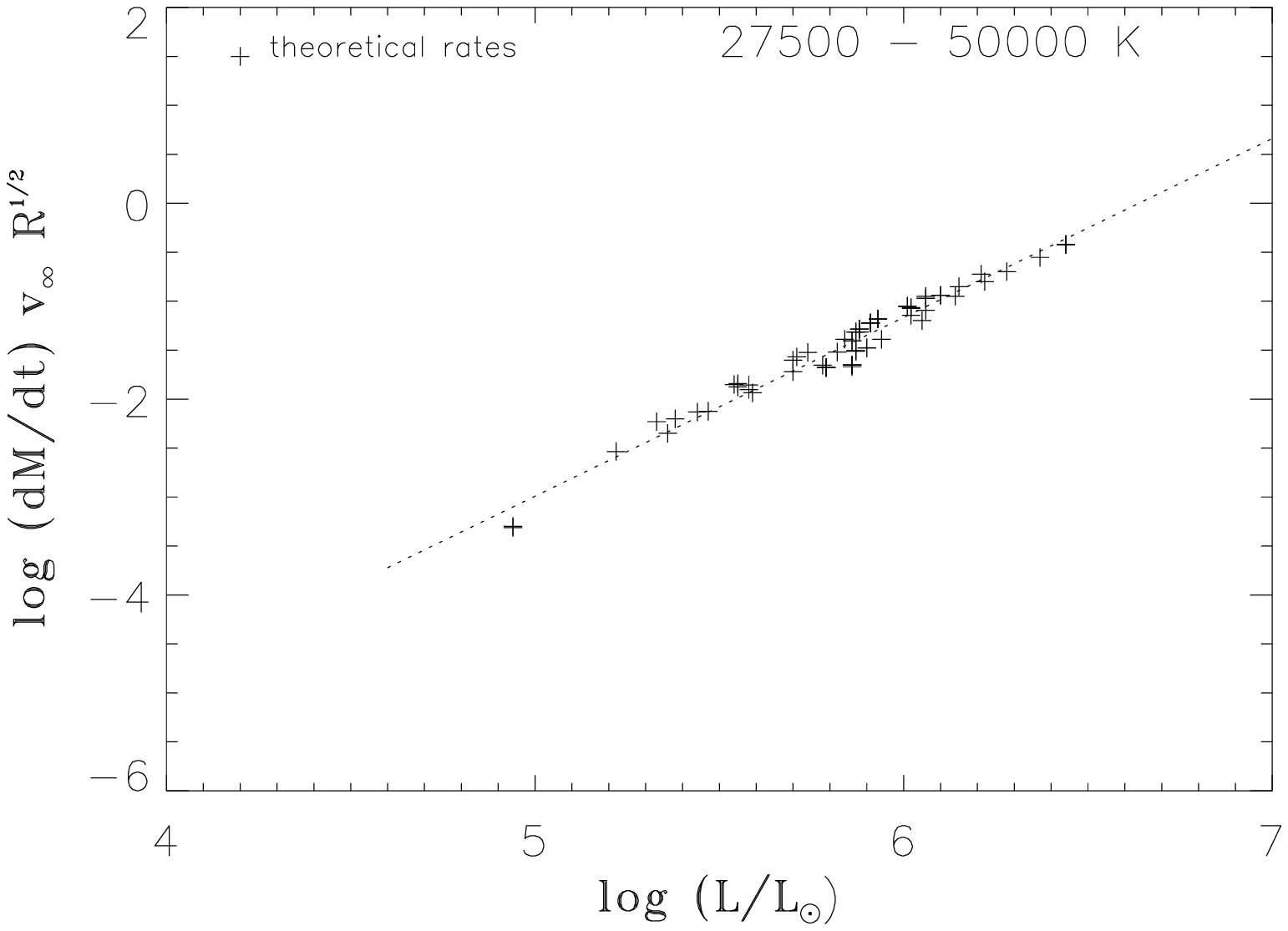, width = 10 cm}}
\centerline{\psfig{file=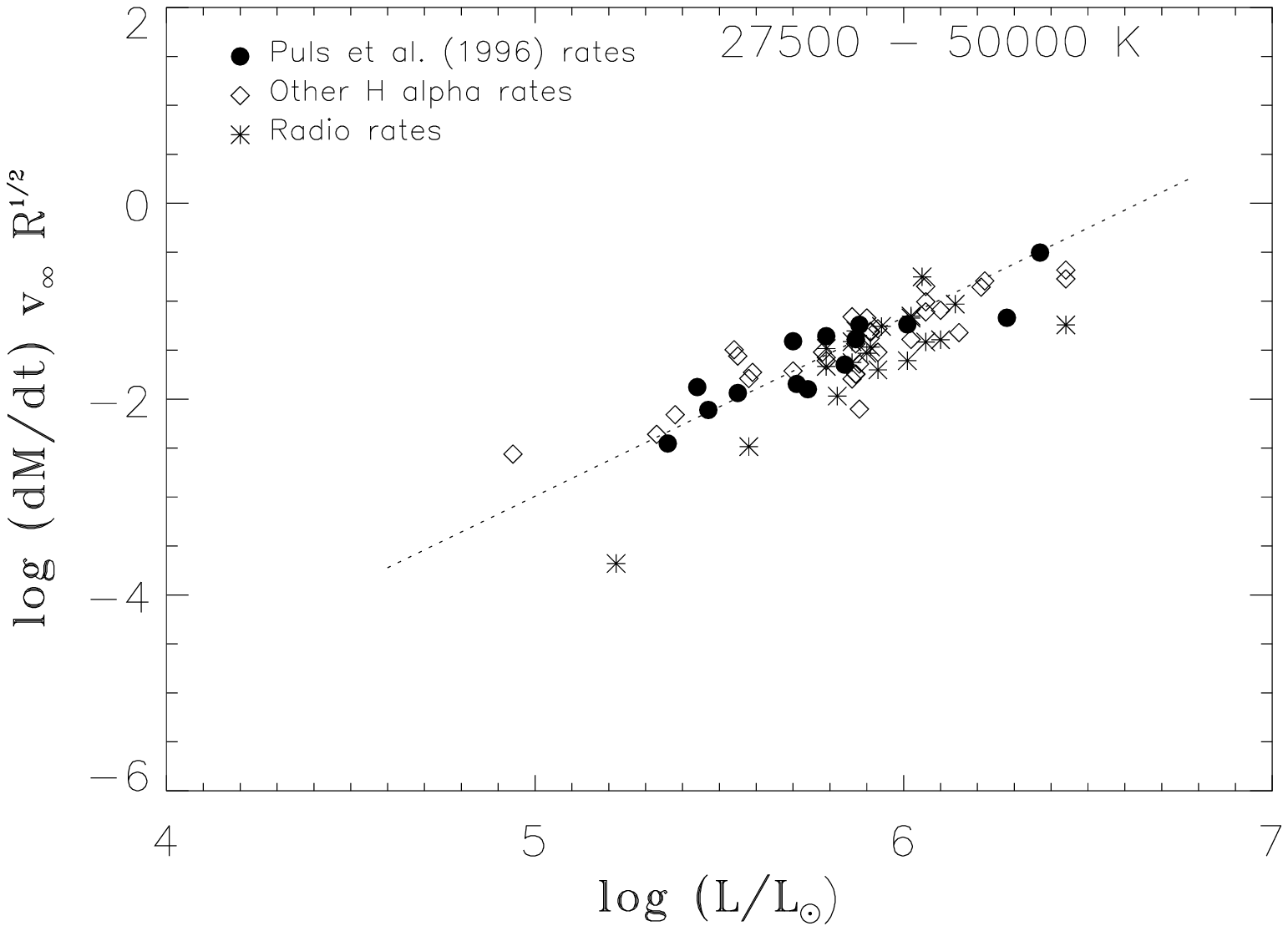, width = 10 cm}}
\caption{Upper panel: The theoretical modified wind momentum
         expressed in $\modfunit$ for the stars in the first $\teff$ range
         (27~500 $< \teff \le$ 50~000 K). 
         The dotted line indicates the best linear fit.
         Lower panel: The observational modified wind momentum for these 
         stars. The dotted line indicates the same theoretical linear fit, 
         as in the upper panel.}
\label{f_modfO}
\end{figure}

We divide the $\teff$ range into two parts. First, we examine
the wind momenta for stars where $\teff \ge$ 27~500 K, later on 
we will also compare the cooler stars.
Figure~\ref{f_modfO} shows the modified wind momentum as a function of 
stellar luminosity for the sample of stars 
with known observational mass-loss rates. The upper panel shows these
modified wind momentum values for the theoretical 
mass-loss rates and a linear best fit
through these theoretical data (dotted line). Note that 
the ``theoretical'' WLR only contains the theoretical $\dot{M}$, 
the included values for $\vinf$ and $R_*$ were taken 
from observations.  
The theoretical WLR is:

\begin{eqnarray}
\Pi^{\rm theory} & = &~-12.12~(\pm~0.26) \nonumber \\
                 & & ~+~1.826~(\pm~0.044)~{\rm log}(L/\lsun) \nonumber\\
                 \nonumber\\
                 & &~{\rm for}~\teff~\ge~27~500~{\rm K}
\label{eq_wlro}
\end{eqnarray}
Since the slope of the WLR of Eq.~(\ref{eq_wlro}) has a slope of $x = 1.826$, the derived 
theoretical value for $\alpha_{\rm eff}$ (Eq.~\ref{eq_alphaeff}) that follows, is: 

\begin{equation}
\alpha_{\rm eff}~=~\frac{1}{x}~=~0.548 \nonumber
\end{equation}
This corresponds well to predicted values of the force multiplier 
parameter ($\alpha \simeq$ 0.66 and $\delta \simeq$ 0.10, see e.g. Pauldrach et al. 1994).

The lower panel of Fig.~\ref{f_modfO} shows that both the 
WLR for the Puls et al. (1996) data and that
for the other methods/authors, follow the same relationship, both in  
agreement with the theoretical WLR. The dotted line is 
again the theoretical best linear fit.
We conclude that for the range of the O stars, there is good agreement between
theoretical wind momenta and those determined from observations.
The scatter between theoretical and observational 
modified wind momenta is only 0.06 (1 $\sigma$). 

The good agreement between the observational and theoretical wind momenta 
adds support to the possibility to derive distances to luminous, hot 
stars in extragalactic stellar systems using the WLR. 
In practice the technique may be hampered by e.g. the fact that 
O stars are mostly seen in stellar clusters and cannot be spatially resolved
in distant stellar systems. 
This is one of the reasons why the visually brighter B-type and 
especially the A-type supergiants located in the field are 
expected to be better candidates in actually using the WLR 
as a distance indicator (see Kudritzki et al. 1999). 

Comparison between
the theoretical and observational WLR for the winds
of B and A type supergiants is thus essential to
investigate whether the slope of the WLR is the same for
different spectral ranges. This is not expected, since
the winds of different spectral types are driven by lines
of different ions (see Vink et al. 1999; Puls et al. 2000).


\subsection{Modified Wind momentum comparison for Range 2 \\
~~~~~(12~500 $\le \teff \le$ 22~500 K)}
\label{s_modfcompB}

\begin{figure}
\centerline{\psfig{file=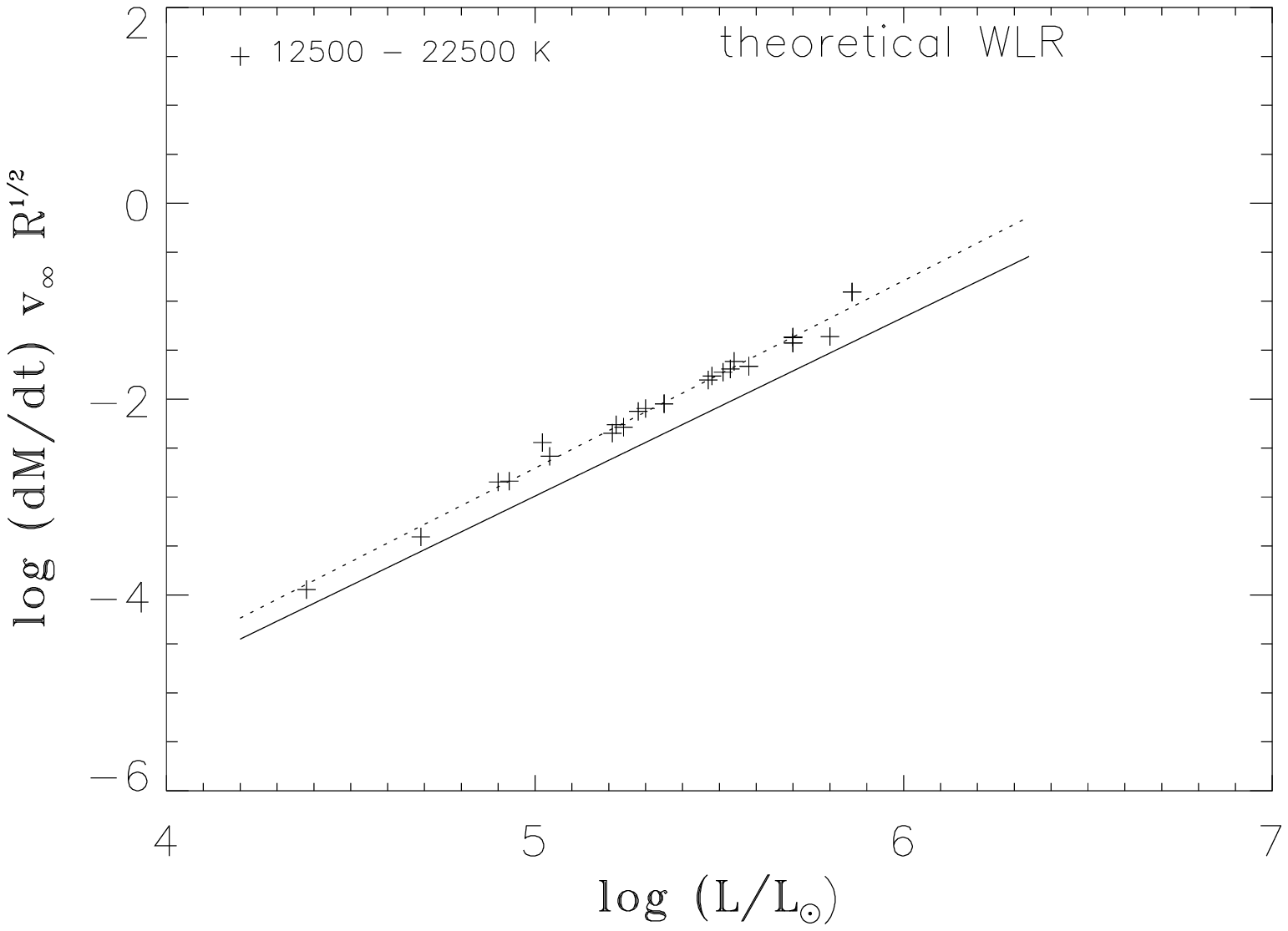, width = 10 cm}}
\centerline{\psfig{file=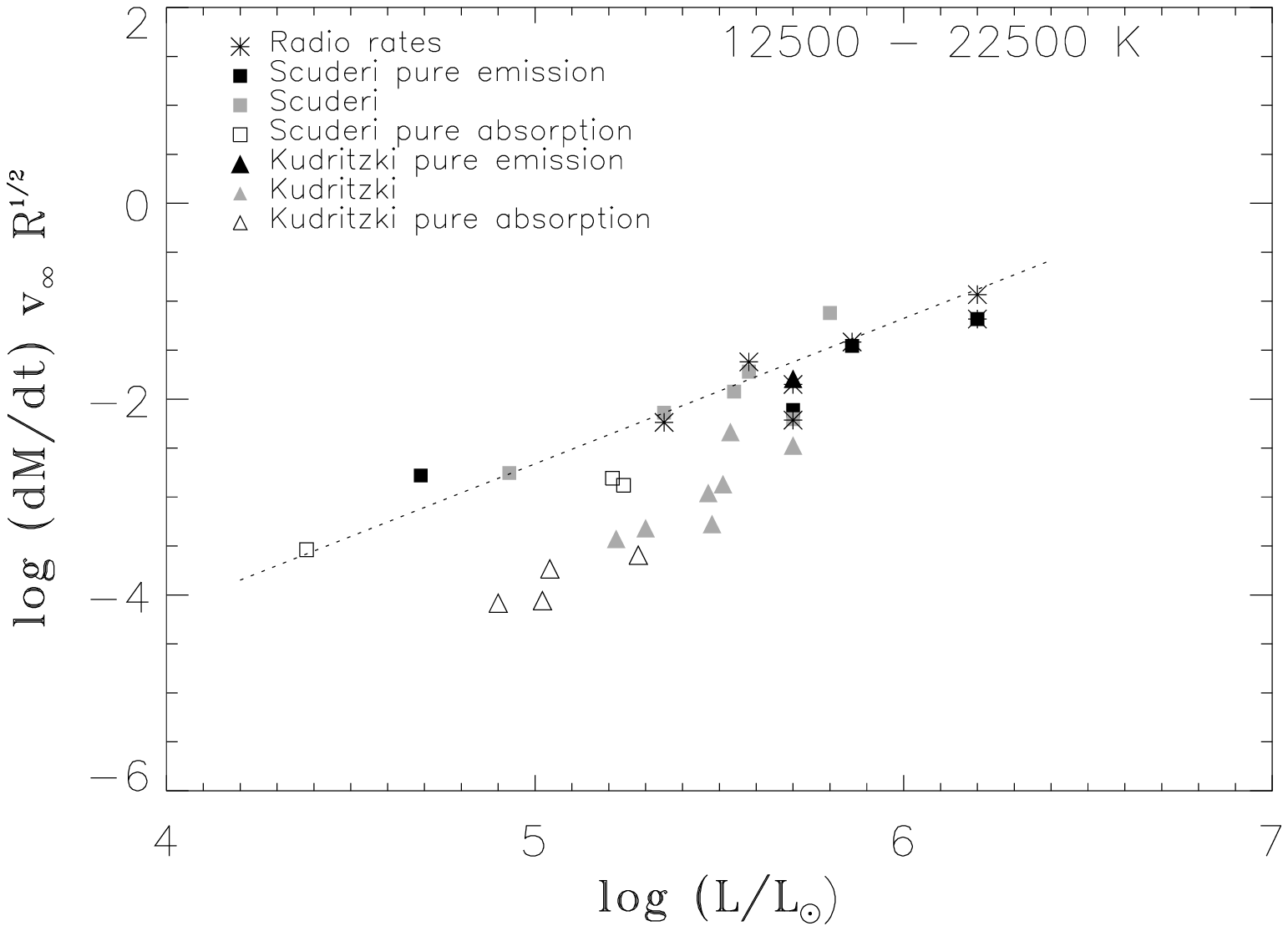, width = 10 cm}}
\caption{Upper panel: The {\it theoretical} modified wind momentum
         expressed in $\modfunit$ for the second range (12~500 $\le \teff \le$ 22~500 K).
         The {\it dotted} line indicates the best linear fit for this range. 
         The solid line indicates the theoretical WLR for stars in the 
         range 27~500 $< \teff \le$ 50~000 K. 
         Lower panel: The {\it observational} modified wind momentum for 
         stars in this $\teff$ range (12~500 $\le \teff \le$ 22~500 K). 
         The different sources of the observations are indicated in the upper 
         left corner.
         The dotted line again indicates the {\it theoretical} linear fit 
         for the stars in the second range (12~500 $\le \teff \le$ 22~500 K).}
\label{f_modfB}
\end{figure}

Figure~\ref{f_modfB} shows the modified wind momentum as a 
function of luminosity for both theory and observations for the 
stars in the second range (12~500 $\le \teff \le$ 22~500 K). 
A best fit through the theoretically
derived WLR is indicated with a dotted line in both panels. 
The theoretical WLR for this $\teff$ range is:

\begin{eqnarray}
\Pi^{\rm theory} & = &~-12.28~(\pm~0.23) \nonumber\\
                 & &~+~1.914~(\pm~0.043)~{\rm log}(L/\lsun) \nonumber\\
                 \nonumber\\
                 & &~{\rm for}~12~500~\le~\teff~\le~22~500~{\rm K}
\end{eqnarray}
Since the slope of the WLR for this range is slightly higher  
than that for the O star range, the 
predicted value for $\alpha_{\rm eff}$ is somewhat lower (see Sect.~\ref{s_mwm}), 
namely: 

\begin{equation}
\alpha_{\rm eff}~=~\frac{1}{x}~=~0.522 \nonumber\\
\end{equation}

The lower panel of Fig.~\ref{f_modfB} indicates the observed modified 
wind momenta (the dotted line contains the {\it theoretical}
mass-loss rates).
For this second $\teff$ range (12~500 $\le \teff \le$ 22~500 K) the
plot in the lower panel reveals a large scatter in the observed data.

Comparison of these observations with our predictions shows that within 
the subset of radio mass-loss rates there does not appear to be a systematic 
discrepancy. Also, those H$\alpha$ profiles which are fully in emission (the 
filled symbols in the lower panel of Fig.~\ref{f_modfB}), i.e. 
the profiles that within the H$\alpha$ method most likely provide the most 
reliable mass-loss rates, do not show a systematic difference with the 
radio rates. The picture becomes different for stars that show H$\alpha$
to be P~Cygni shaped (grey symbols in lower panel of Fig.~\ref{f_modfB}) 
or fully in absorption (open symbols).
Although the measurements of Scuderi (1994,2000) remain reasonably consistent, 
those by Kudritzki et al. (1999) are discrepant in that 
at log $L/\lsun \simeq 5.8$ these values start to diverge from 
the other observed rates, such that below log $L/\lsun \simeq 5.6$ a systematic difference of 
about a factor of 30 results between different sets of observed mass-loss rates.

An investigation of the origin of these systematic differences in 
observed B star wind momenta is beyond the scope of this paper. 
We will address this issue in a separate study (Lamers et al., in preparation). 
Here we just note that the large scatter in the observed H$\alpha$ data implies 
that there is either a dichotomy in the wind-momenta of B-stars (as
suggested by Kudritzki et al. 1999) or that 
there exist systematic errors in the mass-loss determinations from 
H$\alpha$ for B stars.

The systematic discrepancies for the observed B star wind momenta imply that 
we cannot currently compare our predictions with observed data in the most
meaningful way, since the data are not consistent and thus a fair 
comparison with our predictions cannot be conclusive.
In addition, it may be meaningful to further investigate the validity
of assumptions in our method of predicting the mass-loss rates of B-type 
stars (see e.g. Owocki \& Puls 1999).
Still, we note that the most reliable rates (from radio and pure H$\alpha$ 
emission profiles) appear to be consistent with our predictions. 

The upper panel of Fig.~\ref{f_modfB} reveals that most of the
models in the second $\teff$ range (12~500 $\le \teff \le$ 22~500 K) 
lie {\it above} the theoretical fit for the models from the first $\teff$ 
range ($\teff \ge$ 27~500 K). This is due to the increase in the mass-loss 
rate at the bi-stability jump of a factor of five. The models with 
12~500 $\le \teff \le$ 22~500 K are, however, only slightly above 
the fit for the O star models ($\teff \ge$ 27~500 K), as at 
the bi-stability jump the terminal velocity $\vinf$ drops by a 
factor of two.


\section{Discussion}
\label{s_disc}

We have shown that our predictions of mass loss for O stars, using
Monte Carlo simulations of energy loss during photon transport in non-LTE
unified wind models, yields good agreement with the observed values.
This demonstrates that an adequate treatment of ``multiple scattering'' in
radiation-driven wind models resolves the discrepancy between observations
and theory that had been reported earlier. 
The agreement between observed and theoretical wind momenta of
O stars adds support to the method of deriving distances to distant stellar
systems using the WLR.

The comparison between the predicted and observed values 
of the modified wind momentum $\Pi$ for the B stars is not 
conclusive. 
A good comparison between the 
observations and our predictions for the 
B star regime needs to await an explanation of the  
discrepancies in the observed B star mass-loss rates. This issue will 
be addressed in a separate study.

Our models predict a jump in mass loss of about a factor of five around
spectral type B1. An important point that supports this prediction 
is the following. Vink et al. (1999) have calculated the mass-loss rate and $\vinf$
for winds at both sides of the bi-stability jump in a self-consistent way
for models with log(L/$\lsun$) = 5.0 and M = 20~$\msun$.
These self-consistent calculations showed a jump in $\dot{M}$
of a factor of five and a simultaneous drop in $\ratio$ of a factor of 
two. This drop in $\ratio$ has been {\it observed} (Lamers et al. 1995). 
This gives support to our prediction that the mass-loss rate 
at spectral type B1 increases by the predicted amount. 

Since there is good agreement between observed mass-loss rates by different
methods and the new theoretical mass-loss rates for the O-type stars, whereas 
there is inconsistency between the observed mass-loss rates from different 
authors for the B-type stars, this may point to the presence of 
systematic errors in mass-loss determinations from observations for B stars.

Because our predictions for the O stars agree with observations and
our models also predict the bi-stability jump
around spectral type B1, we believe that our theoretical mass-loss predictions 
are reliable and suggest they be used in new evolutionary 
calculations of massive stars.


\section{Summary \& Conclusions}
\label{s_concl}

\begin{enumerate}

\item{} We have calculated a grid of wind models and
mass-loss rates for a wide range of stellar parameters, corresponding to 
masses between 15 and 120 $\msun$.

\item{} We have derived two fitting formulae for the mass-loss
rates in two ranges in $\teff$ at either side of the bi-stability jump
around 25~000 K. A mass-loss recipe was derived that connects the two
fitting formulae at the bi-stability jump.

\item{} There is good agreement between our mass-loss predictions
that take {\it multiple scattering} into account, 
and the observations for the O stars. There is no systematic difference 
between predicted and observed mass-loss rates.

\item{} A comparison between observed and predicted wind momenta of O-type stars 
also shows there is good agreement. This adds support to the use of the WLR as a way
to derive distances to luminous O stars in distant stellar systems.

\item{} For the observed mass-loss rates of B stars there is an inconsistency 
between rates derived by different authors and/or methods. One group
of $\dot{M}$ determinations of B stars does follow the theoretical
relationship, while another group does not. This lack of agreement between 
the observed mass-loss rates of B stars 
may point to systematic errors in the observed values.

\item{} Since our new theoretical mass-loss formalism is successful 
in explaining the observed mass-loss rates for O-type stars, as well as in 
predicting the location (in $\teff$) and size (in $\ratio$) of the 
observed bi-stability jump, we believe that our predictions are 
reliable and suggest that our recipe be used in new evolutionary 
calculations for massive stars. A computer routine to calculate mass 
loss is available either upon request or at the following url: www.astro.uu.nl/$\sim$jvink/.

\end{enumerate}


\begin{acknowledgements}

We thank the referee, Joachim Puls, for constructive comments that
helped improve the paper.
JV acknowledges financial support from the NWO Council for Physical 
Sciences.
AdK acknowledges support from NWO Pionier grant 600-78-333 to L.B.F.M.
Waters and from NWO Spinoza grant 08-0 to E.P.J. van den Heuvel.

\end{acknowledgements}


\begin{thebibliography}{}

\bibitem[Abbott 1982]{abbott82}
       Abbott D.C., 1982, ApJ 259, 282

\bibitem[Abbott \& Lucy 1985]{abbott85}
       Abbott D.C., Lucy L.B., 1985, ApJ 288, 679

\bibitem[Achmad et al. 1997]{achmad97}
       Achmad L., Lamers H.J.G.L.M., Pasquini L., 1997, A\&A 320, 196

\bibitem[Allen 1973]{allen73}
       Allen, C.W., 1973, Astrophysical quantities, Athlone Press

\bibitem[Castor et al. 1975]{castor75}
       Castor J.I., Abbott D.C., Klein R.I., 1975, ApJ 195, 157

\bibitem[Chiosi \& Maeder]{chiosi86}
      Chiosi C., Maeder A., 1986, ARA\&A 24, 329

\bibitem[de Koter et al. 1993]{dekoter93}
       de Koter, A.,Schmutz, W., Lamers, H. J. G. L. M., 1993, A\&A 277, 561

\bibitem[de Koter et al. 1997]{dekoter97}
       de Koter A., Heap S.R., Hubeny I., 1997, ApJ 477, 792

\bibitem[Groenewegen \& Lamers]{groen89}
        Groenewegen M.A.T., Lamers H.J.G.L.M., 1989, A\&AS 79, 359

\bibitem[Haser et al. 1995]{haser95}
        Haser S., Lennon D.J., Kudritzki R.-P., 1995,  A\&A 295, 136

\bibitem[Herrero et al. 1995]{herrero99}
        Herrero A., Puls, J., Villamariz, M.R., 2000, A\&A 354, 193

\bibitem[Kudritzki et al. 1989]{kudritzki89}
       Kudritzki R.-P., Pauldrach A.W.A., Puls J., Abbott D.C., 1989, 
       AAP 219, 205

\bibitem[Kudritzki et al. 1995]{kudritzki95}
       Kudritzki R.-P, Lennon D.J., Puls J., 1995, in: ``Science with the VLT'', eds. Walsh J.R., 
       Danziger I.J., Springer Verlag, p. 246

\bibitem[Kudritzki et al. 1999]{kudritzki99}
       Kudritzki R.-P, Puls J., Lennon D.J., et al., 1999, A\&A 350, 970

\bibitem[Lamers \& Leitherer 1993]{lamers93}
        Lamers H.J.G.L.M., Leitherer, C., 1993, ApJ 412, 771

\bibitem[Lamers et al. 1995]{lamers95}
        Lamers H.J.G.L.M., Snow T.P., Lindholm D.M., 1995, ApJ 455, 269

\bibitem[Lamers et al. 2000]{lamers00}
        Lamers H.J.G.L.M., Nugis T., Vink J.S., de Koter A., 2000, in: ``Thermal and ionization
        aspects from hot stars'', eds. Lamers H.J.G.L.M., Sapar A., ASP Conf Ser 204, p. 395

\bibitem[Leitherer et al. 1995]{leith95}
        Leitherer C., Chapman J., Korabalski B., 1995, ApJ 450, 289

\bibitem[Lucy \& Solomon 1970]{lucy70}
        Lucy L.B., Solomon P., 1970, ApJ 159, 879

\bibitem[Lucy \& Abbott 1993]{lucy93}
        Lucy L.B., Abbott D.C., 1993, ApJ 405, 738

\bibitem[Meynet et al. 1994]{meynet94}
        Meynet G., Maeder A., Schaller G., Schearer D., Charbonel C., 1994, A\&AS 103, 97

\bibitem[Owocki \& Puls 1999]{owocki99}
        Owocki S.P., Puls J., 1999, ApJ 510, 355

\bibitem[Pauldrach et al. 1986]{paul86}
        Pauldrach A.W.A., Puls J., Kudritzki R.P., 1986, A\&A 164, 86

\bibitem[Pauldrach et al. 1994]{paul94}
        Pauldrach A.W.A., Kudritzki R.P., Puls J., Butler K., Hunsinger J.,1994, A\&A 283, 525

\bibitem[Puls]{puls87}
        Puls J., 1987, A\&A 184, 227

\bibitem[Puls et al. 1996]{puls96}
        Puls J., Kudritzki R.P., Herrero A., et al., 1996, A\&A 305, 171

\bibitem[Puls et al. 2000]{puls00}
        Puls J., Springmann U., Lennon M., 2000, A\&AS 141, 23 

\bibitem[Schmutz 1991]{schm91}
        Schmutz W., 1991, in: ``Stellar Atmospheres: Beyond Classical Models'',
        eds. Crivellari L., Hubeny I., Hummer D.G., NATO ASI Series C, Vol. 341, 191

\bibitem[Scuderi 1994]{scud94}
        Scuderi S., 1994, ``Properties of winds of early type stars'', thesis, Univ. of Catania

\bibitem[Scuderi \& Panagia 2000]{scud00}
        Scuderi S., Panagia N., 2000, in: ``Thermal and ionization aspects from hot 
        stars'', eds. Lamers H.J.G.L.M., Sapar A., ASP Conf Ser 204, p. 419

\bibitem[Scuderi et al. 1992]{scud92}
        Scuderi S., Bonanno G., Di Benedetto R., Sparado D., Panagia N., 1992, A\&A 392, 201

\bibitem[Scuderi et al. 1998]{scud98}
        Scuderi S., Panagia N., Stanghellini C., Trigilio C., Umana C., 1998, A\&A 332, 251

\bibitem[Springmann 1994]{spring94}
        Springmann, U., 1994, A\&A 289, 505

\bibitem[Taresch et al. 1997]{taresch97}
        Taresch, G., Kudritzki, R.P., Hurwitz, M., et al., 1997, A\&A 321,531

\bibitem[Vink et al. 1999]{vink99}
        Vink, J.S., de Koter A., Lamers H.J.G.L.M., 1999, A\&A 350, 181


\end{thebibliography}
\end{document}